\RequirePackage{fix-cm}

\documentclass{svjour3}        
\smartqed  
\usepackage{graphicx}
\usepackage{latexsym}
\usepackage{mathrsfs}
\usepackage{amssymb,amsbsy}
\usepackage{amsmath}
\usepackage{amsfonts}
\usepackage{natbib}
\newcommand{\apj}{\it Astrophys.~J.}
\newcommand{\aap}{\it Astronom.~Astrophys.}
\newcommand{\apss}{\it Astrophys. Space Sci.}

\newcommand{\aj}{\it Astronom.~J.}
\newcommand{\apjl}{\it Astrophys.~J.~Lett.}
\newcommand{\apjs}{\it Astrophys.~J.~Suppl.~Ser.}

\newcommand{\mnras}{\it Mon.~Not.~R.~Astron.~Soc.}
\newcommand{\nature}{\it Nature}

\newcommand{\grl}{\it Geophys.~Res.~Lett.}
\newcommand{\as}{\prime\prime}
\def\deg{\circ}

\newcommand{\g}{\gamma}

\newcommand{\D}{\Delta}

\renewcommand{\O}{\Omega}

\def\be{\begin{equation}}
\def\ee{\end{equation}}

\def\fr{\frac}

\journalname{}

\begin{document}

\title{Prospect for UV observations from the Moon}

\titlerunning{Lunar UV telescope}        

\author{Margarita Safonova, Joice Mathew, Rekhesh Mohan, A.~G.~Sreejith, Jayant Murthy,
Noah Brosch, Norbert Kappelmann, Arpit Sharma and Rahul Narayan}

\authorrunning{Safonova, et al.} 
\institute{Margarita Safonova
           \and
Joice Mathew
           \and
Rekhesh Mohan
		\and
A.~G.~Sreejith
           \and
Jayant Murthy\at
Indian Institute of Astrophysics, Koramangala 2nd block, Bangalore, 560034, India \\
              \email{rita@iiap.res.in}        \\
           \and
Noah Brosch\at
The Wise Observatory and the Dept. Of Physics and Astronomy, Tel Aviv University, Tel Aviv 69978, Israel \\
            \and
Norbert Kappelmann\at
Kepler Center for Astro and Particle Physics, Institute of Astronomy and
Astrophysics, University of T\"{u}bingen, Sand 1, 72076, T\"{u}bingen, Germany\\
           \and
Arpit Sharma          
          \and
Rahul Narayan\at
Team Indus, Axiom Research Labs Private Limited, D-37, Sector-63, Noida, UP, India}

\date{Received: date / Accepted: date}

\maketitle

\begin{abstract}

Space astronomy in the last 40 years has largely been done from spacecraft in low Earth orbit (LEO) 
for which the technology is proven and delivery mechanisms are readily available. However, new opportunities 
are arising with the surge in commercial aerospace missions. We describe here one such possibility: 
deploying a small instrument on the Moon. This can be accomplished by flying onboard the Indian 
entry to the Google Lunar X~PRIZE competition, Team Indus mission, which is expected to deliver a nearly 30 kgs 
of payloads to the Moon, with a rover as its primary payload. We propose to mount a wide-field far-UV (130--180 nm) 
imaging telescope as a payload on the Team Indus lander. Our baseline operation is a fixed zenith pointing but 
with the option of a mechanism to allow observations of different attitudes. Pointing towards intermediate ecliptic 
latitude ($50^{\deg}$ or above) ensures that the Sun is at least $40^{\deg}$ off the line of sight at all times. 
In this position, the telescope can cover higher galactic latitudes as well as parts of Galactic plane. The scientific 
objectives of such a prospective are delineated and discussed.

\end{abstract}

\section{Introduction}
\label{intro}

The Earth's atmosphere absorbs and scatters UV photons preventing observations of the 
active Universe. UV-emitting phenomena are generally associated with high-energy 
activity: massive star formation, hot transients such 
as supernovae (SNe), which are UV bright for hours to days, variability of AGN, where 
the amplitude of the variability usually increases at shorter wavelengths, variable stars such as 
M-dwarfs with UV-flaring activity on time scales of hundreds of seconds, and flashes from 
cosmic collisions which can be very energetic on all scales. The UV range is a critical tool for 
classifying and studying these hot transients. 

Tidal disruption events (TDEs) by massive black holes give rise to yet another source of
flares in the UV, where the object can continue to be UV bright for months. A supermassive 
black hole (SMBH) at the centre of a typical galaxy disrupts a main sequence (MS) star 
every $10^4$--$10^5$ years (e.g. Alexander 2005; Gezari et al. 2009), and a planet 
as often as that or even more considering that the expected number of free-floating 
planets exceeds the number of stars by at least two orders of magnitude \citep{Gahm2013}. 

More than a thousand exoplanets are now known (about 1800 planets confirmed 
and more than 3800 waiting for confirmation at time of writing\footnote{Extrasolar Planets Encyclopedia, 
{\tt http://http://exoplanet.eu/catalog/}, June 2014} with more than a third of them being young systems with 
host ages of less than 3 Gyr (Schekinov et al. 2014). In such systems collisions at all scales are possible: 
planet-planet, planet-star and collisions of smaller objects/debris (planetesimals, 
asteroids, comets) which result in flashes and flares peaking in the extreme-UV (EUV)/UV. 
Even in our own Solar System, there are a large number of asteroidal collisions which
create rock fragments of all sizes that subsequently impact planets. For example, an impact on 
the asteroid Vesta has been identified as the most probable cause of HED meteorites (Howardite, 
Eucrite, Diogenite class of 
achondrites) influencing the near-Earth space (see Pieters et al. 2006 and references therein). 
Recent cataclysmic events such 
as the Shoemaker-Levy 9 (SL-9) comet impact on Jupiter and the 2013 Chelyabinsk meteor crash
in Russia (the most probable parent of this $\sim 10,000$ tons and 18-m diameter bolide 
is the shattered PHA 2011 EO40, which has a possible family of up to 20 asteroids of different sizes
(de la Fuente Marcos \& de la Fuente Marcos 2012), have renewed attention to such threats.

\begin{figure}[hb!]
\centering{
\includegraphics[scale=0.5]{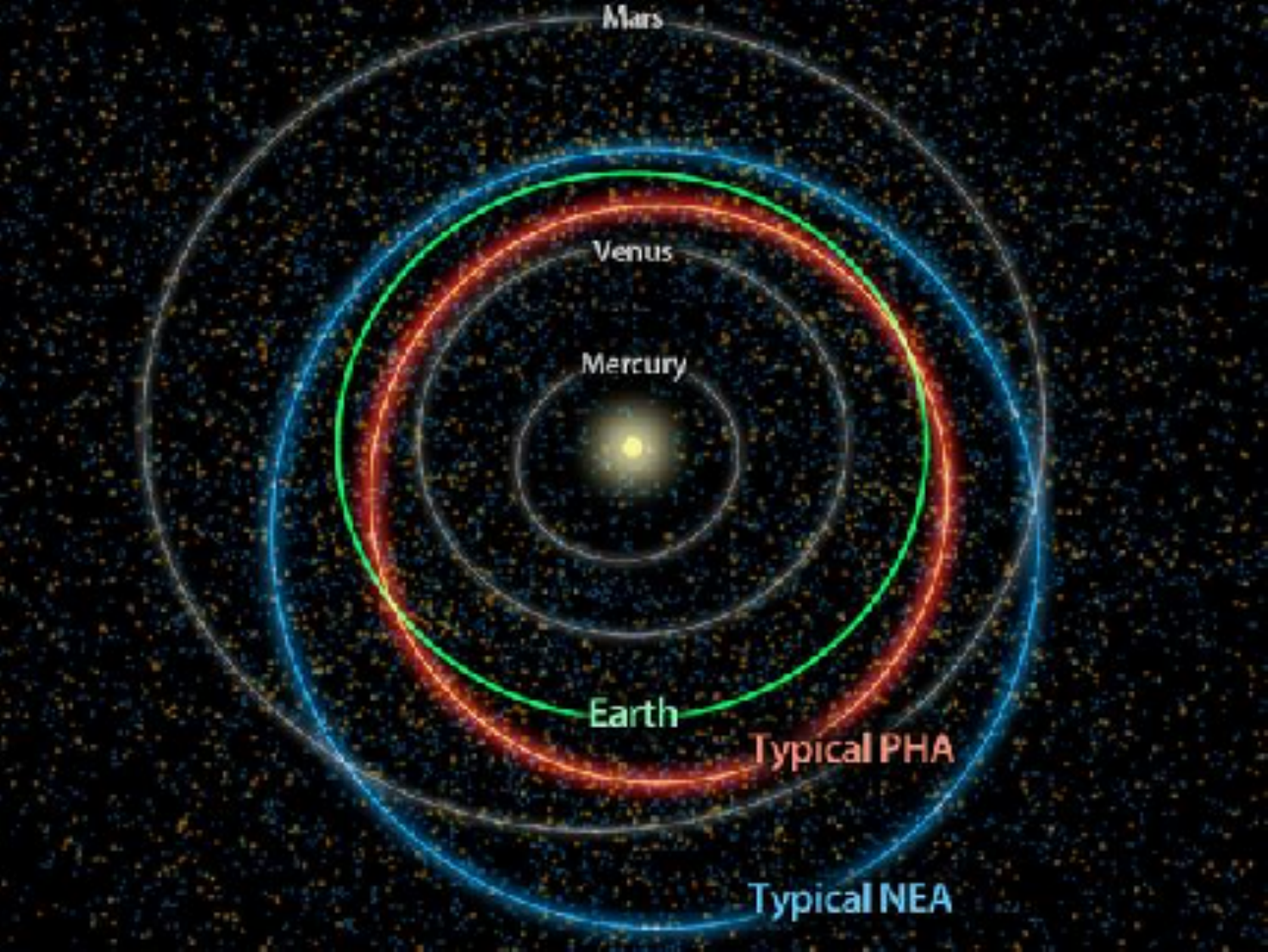}}
\caption{Illustration of orbits of NEO and PHA. Source: NASA}
\label{fig:1}
\end{figure}

Earth is constantly bombarded from space. Meteoroids of size from tens of cms and 
larger enter the Earth's atmosphere on daily to monthly basis. If they are smaller 
than some tens of metres, they ablate before reaching the ground. An object with a mass 
of 10,000 kg collides with Earth approximately every 1--5 years. Though at
present there are approximately 620,000 asteroids that are tracked in our Solar System, this
number is less than one percent of the estimated objects that orbit the Sun and 
less than $10\%$ of all Near-Earth Objects (NEO) have been discovered (Yeomans 2013). 
Moreover, while NEO-participating observatories look for them in the
Northern Hemisphere (such as e.g., Palomar Transient Factory, the Catalina Sky Survey, Pan-STARRS
and SkyMapper), not much is done in the South (the only one observatory in the Southern Hemisphere -
at Siding Spring, Australia - has reduced its operations to only occasional observations
due to budget cuts, and the Large Synoptic Survey Telescope (LSST) is still some years away 
from operation). The best way to find possible impactors --- potentially hazardous asteroids 
(PHA)\footnote{these are the asteroids which can come within 0.05 AU from the Earth and which 
are larger than 140 m in size.} (Fig.~\ref{fig:1}) --- would be to put a space wide-field 
near-IR telescope to look outward from the orbit of Venus.

But the second best, as compared to the more affordable LEO space telescopes, 
is the Moon. This would allow observing the incoming objects on their last part of the trajectory 
before impacting. Although not useful for early warning, this information would be helpful in 
understanding the early stage of the interaction of an incoming body with the outermost layers 
of the atmosphere.

The Moon was first proposed by astronomers as a prime site for large telescopes nearly 50 
years ago (Tifft 1966). Essential absence of atmosphere and ionosphere offered an unobstructed 
view of the space in all wavelengths, and lunar surface and low gravity offered a stable 
platform on which to build large telescopes. The Moon has several advantages
and disadvantages as a base for astronomical observatories, thus scientists cannot 
come to a consensus about the benefits of using the Moon as a site for deep space 
astronomy (e.g., Brosch 2009). On the one hand, in the near-vacuum of the Moon, there 
are no atmospheric related problems like opacity, thermal conductivity, wind, and seeing. 
Observations can be made continuously, provided that a telescope is shielded from
the Sun and has constant power supply. The lunar regolith is fairly stable and offers 
a firm surface on which to build and operate telescopes without requiring costly and 
fault-prone stabilizing mechanisms. On the other hand, the successful operation of the Hubble Space
Telescope (HST) has shown that space-based telescopes have fewer problems in the much more 
easily accessible LEO environment. There is also a possibility of regular servicing, which is 
harder to do on the Moon. However, the trend of placing astronomical telescopes in the $L_2$  
Lagrangian point indicates that the issue of servicing is minor, in comparison to the quality of 
the location. Therefore it was decided that space observatories are more cost-effective 
than lunar-based observatories. 
However, the ambitions were always for the large lunar telescopic installations with 4--100 m 
mirrors or total mass of 10--30 tons of radio antennas (e.g. Douglas \& Smith 1985; Angel et al. 2008), 
driven possibly by the fact that since we are going to the Moon anyway, we can as well make it 
big\footnote{as the the famous quote by Kraft Ehricke goes ``The exploration and settlement of the
Moon must be done on a grand scale" (Benaroya 2010).}.
 
A UV/optical transit telescope that uses the slow lunar sidereal rate was first proposed by 
McGraw (1994), with further development by (Nein \& Hilchey 1995). Modest-sized UV/O/IR 
fully-steerable robotic telescopes using lightweight optics have been also proposed \citep{Chen1995}
and it was suggested that a large 20-m liquid mirror telescope (Angel et al. 2008) with UV 
imaging capability has the potential to surpass the sensitivity of HST and even JWST by orders of 
magnitude (Klimas et al. 2010). However, the first and only astronomical UV observations from the 
Moon to date were made by Apollo 16 team in 1972.

We were recently given an opportunity to fly an astronomical instrument onboard the 
Team Indus Lunar lander to the Moon. Team Indus\footnote{Official website is http://blog.teamindus.in} 
is an Indian entry to the Google Lunar X~PRIZE (GLXP) competition\footnote{The Google Lunar X~PRIZE 
is the international \$30 million competition to run a rover for at least 500 m on the lunar surface, 
sponsored by Google and operated by the X~PRIZE Foundation; official website 
{\tt http://www.googlelunarxprize.org/}.}. This mission is a soft-landing mission and carries a 
mini-exploration rover amongst other instruments. After the touchdown, all scientific instruments 
will be turned on in a pre-defined sequence with the highest priority being accorded to the Team 
Indus' rover deployment.

Our primary challenge was to come up with an instrument that would do good science within the 
constraints imposed by the primary mission, one of the major is that we have to deliver the 
payload by mid-2015. We have chosen a wide-field UV telescope --- Lunar Ultraviolet Cosmic 
Imager (LUCI), operating in the far-UV (FUV) range of 130--180 nm, taking advantage of 
the transparency of the lunar sky to the UV.
Our primary science goals is detection of transients such as, for example, TDEs, or SNe in 
distant galaxies as a probe for cosmological distant scale. SNe discovered in optical surveys,
are usually caught too late, when the UV emission is already fading. If we detect SNe early, 
the hot thermal emission from the ejecta is bright in the UV. As we perform
our survey of the sky, we will also pick up other transients such as near-Earth asteroids,
as well as produce an FUV catalog of the sky. We will also map the hot stellar distribution in 
the Galactic disk.

\section{History of astronomical observations from the Moon}
\label{sec:1}

\subsection{Previous science from the Moon}
\label{sec:2}

The very first instrument to fly to the Moon was Luna 1, a Soviet lunar impactor which,
unfortunately,  
missed the Moon in January 1959 and became a solar orbiter. It performed the first-ever direct
measurements of the solar wind, measured the particle level in the outer radiation
belts and found no signature of a lunar magnetic field. Actual scientific exploration of
and from the Moon began when Luna 2 landed on the Moon on September 14, 1959.
The instruments were similar to those of Luna 1. Luna 2 detected fluctuations 
in the electron flux and energy 
spectrum in the Van Allen radiation belt, finding no evidence of a lunar magnetic field or 
radiation belt. Further spacecraft in the Luna series continued the exploration of
the Moon with Luna 3 taking the first photographs of the far side of the Moon on October 7 1959,
showing far fewer signs of volcanic plains than expected which stimulated a revision of lunar evolution 
theories. Subsequently, Luna 9 and 10 measured the radiation fields in 
near-lunar space and determined that the lunar dusty ground could support a lander.

On August 23 1966, NASA's Lunar Orbiter 1 sent back the world's first image of the Earth from
the vicinity of the Moon. NASA's Ranger missions in the early 1960s, in particular Ranger 7, 8 and 9, 
transmitted high-resolution images of the lunar surface before crashing into it. Following the Rangers,
 the Lunar Surveyor series (1 to 7) at the end of that decade soft-landed television cameras and other 
scientific tools to test the lunar surface prior to human landings in the Apollo program. 
The Apollo missions of 1969--1972 extended the studies of the Earth-Moon
system interactions. 

Lunar exploration was put on halt in 1976 with the completion of Soviet Luna program until 
the Hiten Flyby and Orbiter mission in 1990 \citep{Uesugi1991}, followed by several missions of 
different countries, roughly one every four years. However, these missions were all aimed at studying 
the Moon itself and its environment, rather than using the lunar surface as a base for astronomical observations,
perhaps because of the success of orbital observatories at a fraction of the cost of a lunar base.

\subsection{UV from the Moon}

\begin{figure}[ht!]
\centering{
\includegraphics[scale=0.5]{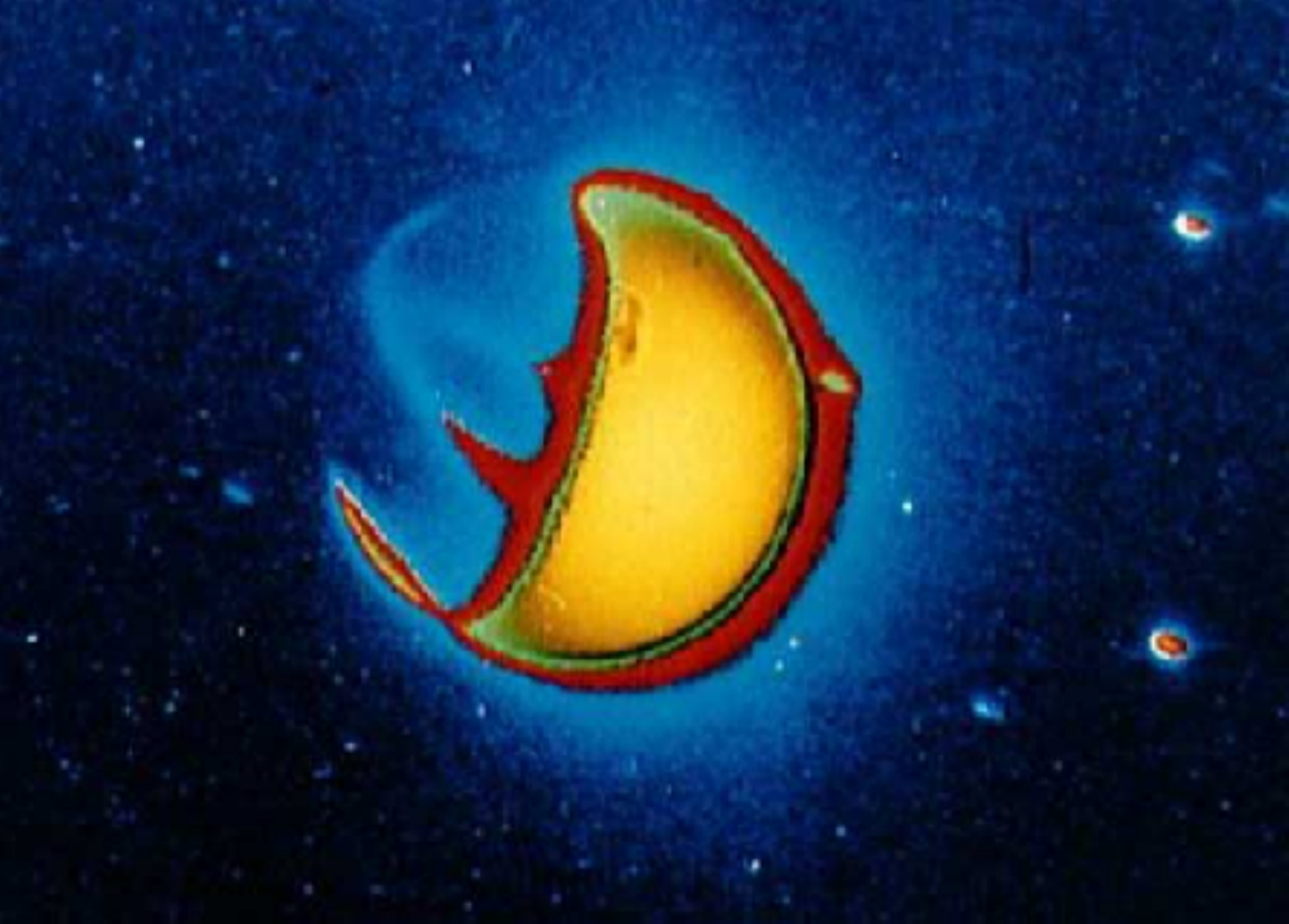}}
\caption{The S201-experiment 10-min exposure photograph in false colours to enhance 
differences in the brightness of the oxygen glow at different layers of atmosphere. For  
this photograph hydrogen emission was blocked so that only oxygen
and nitrogen emissions were recorded. Oxygen atoms fluoresce under the solar UV radiation, and 
maximum concentration is shown in golden 
closest to the Earth's surface. As it reduces with altitude, the colours become green and 
finally blue. On the Earth's night side the gas shows blue in two intersecting arcs over 
the equator --- the so-called equatorial FUV arcs (Hicks \& Chubb 1970). In the South 
the polar aurora is seen. For the colour version of this figure please see the online publication.
The photographic number of the original black\&white UV camera 
photograph is AS16-123-19657. Source: NASA.}
\label{fig:appollo}
\end{figure}

The S201-experiment FUV camera (50--160 nm) deployed by the Apollo 16 mission is the only UV 
telescope that operated on the surface of the Moon till now (at time of writing).
It had an FOV of 20 degrees and a resolution of 2--4 arcmin (Brosch 2009). It had low sensitivity and 
could only see stars brighter than $V=11$ magnitude but still obtained 
the first FUV atlas of the Large Magellanic Cloud (LMC), as well as of several fields at low 
and high galactic latitudes \citep{{PageCarruthers},{CarruthersPage}}. It also produced the first ever 
images of the Earth in FUV (Fig.~\ref{fig:appollo}). 

Several new missions are now planned and some have already been carried out, e.g. the Chinese Chang’e 3 mission 
launched in 
December 2013, with lander and rover having already survived for several lunar days. The lander is equipped with a near-UV 
telescope (LUT) and $15^{\circ}$-wide Extreme Ultraviolet camera (EUVC) to study Earth's plasmasphere 
(Cao et al. 2011). The lander's panoramic cameras and the EUVC have returned the first optical and UV images of the 
Earth from the Moon’s surface in 4 decades. The Lunar UV-Telescope also completed commissioning operations and is 
expected to perform science operations throughout the lander's projected one-year mission, looking at bright astronomical 
objects, including galaxies, active galactic nuclei (AGN) and bright stars.

\subsection{New Opportunities}

It is quite possible that very soon a floodgate of lunar missions will open --- the unprecedented
opportunities for intensive missions for astronomy on/from the Moon due to the surge in privately 
funded space exploration and through the newly-developing partnership between private companies 
and governments renewed the concept of the Moon as the next place to go. Year 2013 witnessed 
two successful missions, NASA's orbiter LADDEE to study lunar atmosphere, and a China's
Chang’e-3 lander --- a partnership between China and the International Lunar Observatory Association (ILOA).
Several GLXP competition teams could land on Moon in 2015, 
and India, Russia and USA planned missions before 2020, with UK, ESA and 
South Korea to follow after 2020. 

A report on a market study by a London Economics commissioned by 
the X~Prize Foundation estimated that commercial opportunities on the Moon will be worth billions 
of dollars in the next few decades \citep{Barton2013}. One of the private companies, Moon 
Express, is planning to perform astronomical observations from the Moon, developing an 
International Lunar Observatory with a 2-kg optical instrument (ILO-X) to fly aboard Moon Express mission 
as part of the GLXP competition, which will be a precursor to a permanent lunar radio observatory ILO-1, 
a 2-m rigid dish antenna, scheduled for 2018.

It doesn't stop at the Moon --- a new space race seems to have started, a race
between governmental agencies and private firms, and it is may be that the firms will lead. 
The prospecting of deep-space resources have required till now an expensive technology with
funds available only to developed nations' governments, but are quickly becoming cheap enough 
for a private industry or a corporation. The best example is the Google Lunar X~PRIZE endeavour. 

Moreover, a flood of private space-exploratory proposals is coming, and they do provide 
space for science as well. Take, for example, Planetary Resources, Inc., a company founded in 
2009 by Eric C. Anderson and Peter H. Diamandis. The primary business of Planetary 
Resources is defined as prospecting and mining asteroids
with high concentrations of water and precious metals, but they are also collaborating with 
NASA in detecting and tracking NEOs, and readying to launch the Arkyd 100 as the first private low-cost 
space telescope, with follow-up of several more Arkyd 200s (Lewicki et al. 2013). Deep Space 
Industries is preparing a series of low-cost robotic fly-by and rendezvous spacecrafts, 
based on cubesat technologies, to gather data on small PHA-type asteroids (Gump 2013). 
They currently contemplate a round-trip craft that will be able to return 25 to 50 kg of samples 
to near-Earth space, or even directly to Earth. 

Not only industries are involved: the nonprofit charitable organization B612 Foundation is 
building an IR imaging space telescope {\it Sentinel} to launch in 2018 into a Venus-like 
orbit for NEO cataloging, with a proposed mission lifetime of more than 6 years (Lu et al. 2013). 
We live in exciting times indeed and hope that there will be a market for niche scientific payloads
in these missions!

\section{Moon's orbital dynamics --- motion of the lunar sky}

Because of the constraints of time and the available platform, we will not have 3-axis
pointing freedom for LUCI. Therefore we have to understand the motion of the lunar sky
and how our initial orientation determines which sources we will be able to observe. 
The motion of the lunar sky is a result of three major movements: the Moon's axial 
rotation, its revolution around the Earth and the Moon-Earth's system revolution around 
the Sun. There are additional minor contributors, such as the ellipticity of the lunar orbit
around the Earth and its variation, inclination of the lunar orbit to the ecliptic, the 
regression of the lunar nodes with the 18.6 yrs period, precession of apsides (8.85 yrs), the 
precession of the lunar rotational axis to the ecliptic with the period of 18.6 yrs, and the 
motion in space together with the Sun, most of which, however, are only important for 
long-term missions.

\paragraph{Ellipticity}

The Moon's orbit around the Earth is slightly elliptical with an average eccentricity 
of $e=0.05$, which in addition varies from 0.04 to 0.07 with a period of 8.9 years. 
The main observational result of this is the $\sim 5\%$ variation in the size of 
the Earth as seen from the Moon, from $1.8^{\circ}$ in aposelene to $2.0^{\circ}$ 
in periselene.

\paragraph{Tidal locking}

The Moon is tidally locked with the Earth, its sidereal day (27 1/3 terrestrial days) 
is equal to the period 
of its revolution around the Earth. The axis of rotation is nearly perpendicular to 
the orbital plane
at $88^{\circ}28'$. A solar day on the Moon is slightly longer than a sidereal 
day and is equal to the 
synodic month, 29 1/2 terrestrial days and, since the tilt of the lunar axis to 
its solar orbit is nearly
$0^{\circ}$, the Sun traces the same path through the lunar sky all year. 
However, due to libration, 
the Earth, as seen from the Moon, will continuously shift its position on the lunar sky.

\paragraph{Libration} 

The sky from the Moon experiences libration both in latitude and in longitude
which cause celestial sources to shift their coordinate positions on the sky. Librations 
in longitude are due to the non-uniform motion of the Moon along its orbit; the maximal value reaches 
$7^{\circ}45'$. Librations in latitude are due to the $\sim 6.7^{\deg}$ tilt of the lunar axis to 
its orbit around the Earth and have a range of $\pm 7^{\circ}41$. The main observational 
result is that the Earth traces an approximate oval $\sim 18^{\circ}$ diameter on the lunar 
sky in one month.

\paragraph{Length of the day and night---illumination, temperatures}

The solar day on the Moon lasts nearly one terrestrial month, and day and night about 
two terrestrial weeks each. Due to power constraints (Sec.~\ref{sec:power}), we can 
only observe during the lunar daytime, which still gives us nearly two weeks of continuous 
observations, although the Sun has to be avoided. In the case of zenith pointing, 
the Sun will be always more than $40^{\circ}$ away from the pointing direction. 
In case we have the orientation freedom, we will take care to choose pointings that
are a minimum of $40^{\circ}$ away from the Sun. Scattering of sunlight 
from the lunar surface is less of a problem because of the low albedo of the 
lunar surface in the UV (at least three times lower than the Earth's) and the solar flux in 
the FUV spectral range is strongly reduced. Note that 
our spectral range has been chosen specifically to be solar-blind.

At night on the Moon, the temperature of a radiatively cooled structure goes down to 60 K; we
however are not planning to operate at night because there will be no power. During the daytime, the subsolar 
point can reach $T\sim 350$ K, with an average of $\sim 200$ K. However, thermal stresses in the 
telescope would be much reduced compared to a space telescope as the Sun moves slowly on the 
lunar sky, and the illumination is more or less uniform across the telescope structure, 
whereas a space telescope is subject to a large temperature gradient. The lunar
environment is thus strongly favoured if we would be to provide suitable shielding. 
For discussion on the topic of shielding, see Sec.~\ref{sec:shielding}. 

\paragraph{Scanning transit rate---the possibility of deep exposures}

Due to the long synodic month, the motion of the lunar sky is slow and the angular transit of an
object on the horizon is $0.53^{\as}$/sec, instead of $15^{\as}$/sec on the Earth. Other intrinsic 
motions result in maximal gradients of $0.023^{\as}/1000$ sec. in RA and $0.0096^{\as}/1000$ sec. 
in Dec, much less than proposed pixel size (see Table~\ref{table:overview}). It is
therefore possible to obtain deep exposures during a single transit: a point source will cross
a $3^{\circ}$ FOV in $\sim 5.7$ hrs. Since the UV window planned for LUCI is not reachable on 
ground based telescopes, it makes LUCI a unique instrument in this wavelength range. 
A typical sampling is $\sim 5.3^{\as}$/pixel with a total field of view of 3 degrees.

\paragraph{Horizontal coordinate system}

The combination of all lunar motions cause dramatic changes in the positions of the celestial sources
in the lunar sky. Usage of the equatorial coordinate system in description of the positions of the celestial
sources is much more difficult than on Earth, therefore, we will use the lunar horizontal coordinate system
with horizontal coordinates azimuth $A_{\rm L}$, altitude $h_{\rm L}$ and the lunar hour angle 
$\theta_{\rm L}$. However, all the formulae relating lunar $A_{\rm L}, h_{\rm L}, \theta_{\rm L}$ to 
the lunar equatorial coordinates of a celestial source have the same form as on Earth.  
We will transform the known celestial coordinates of program sources into the lunar 
horizontal coordinates for a given lunar sidereal time at the selenographic coordinates of the lander.

\section{LUCI --- an FUV wide-field imager from a lunar platform}

\subsection{Optical design, mirrors and structure}

\begin{figure}[hb!]
\includegraphics[scale=0.18]{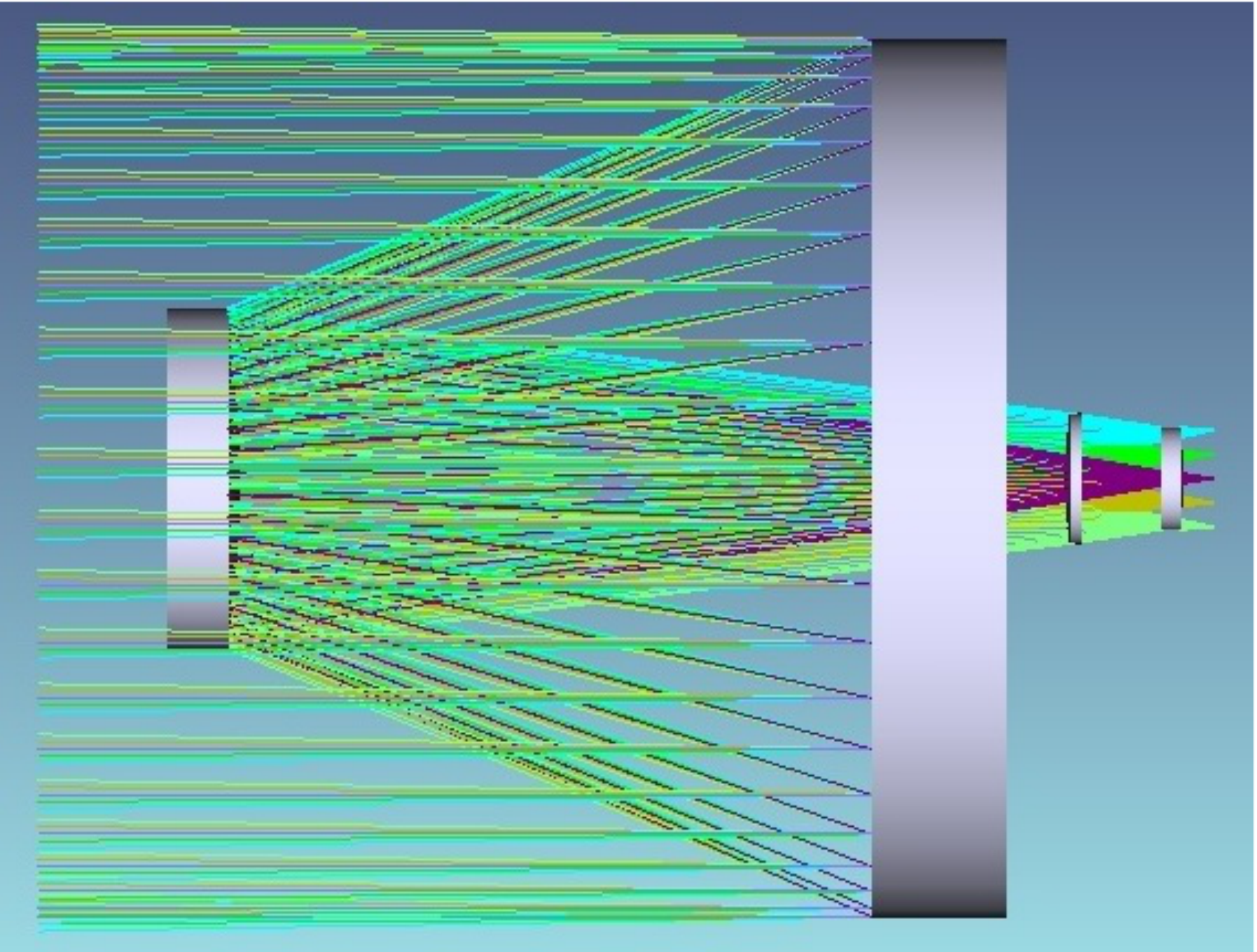}
\includegraphics[scale=0.2]{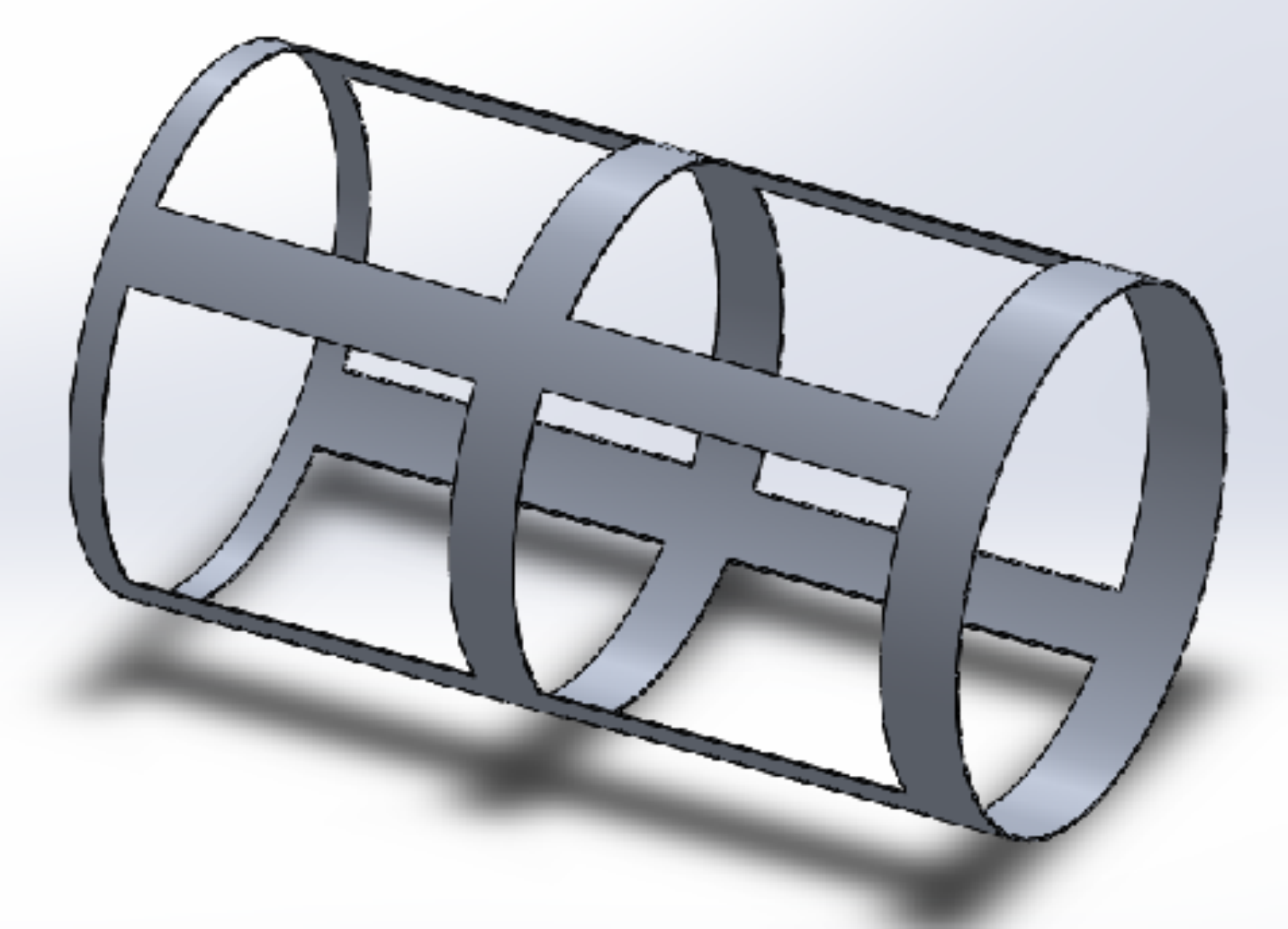}
\includegraphics[scale=0.22]{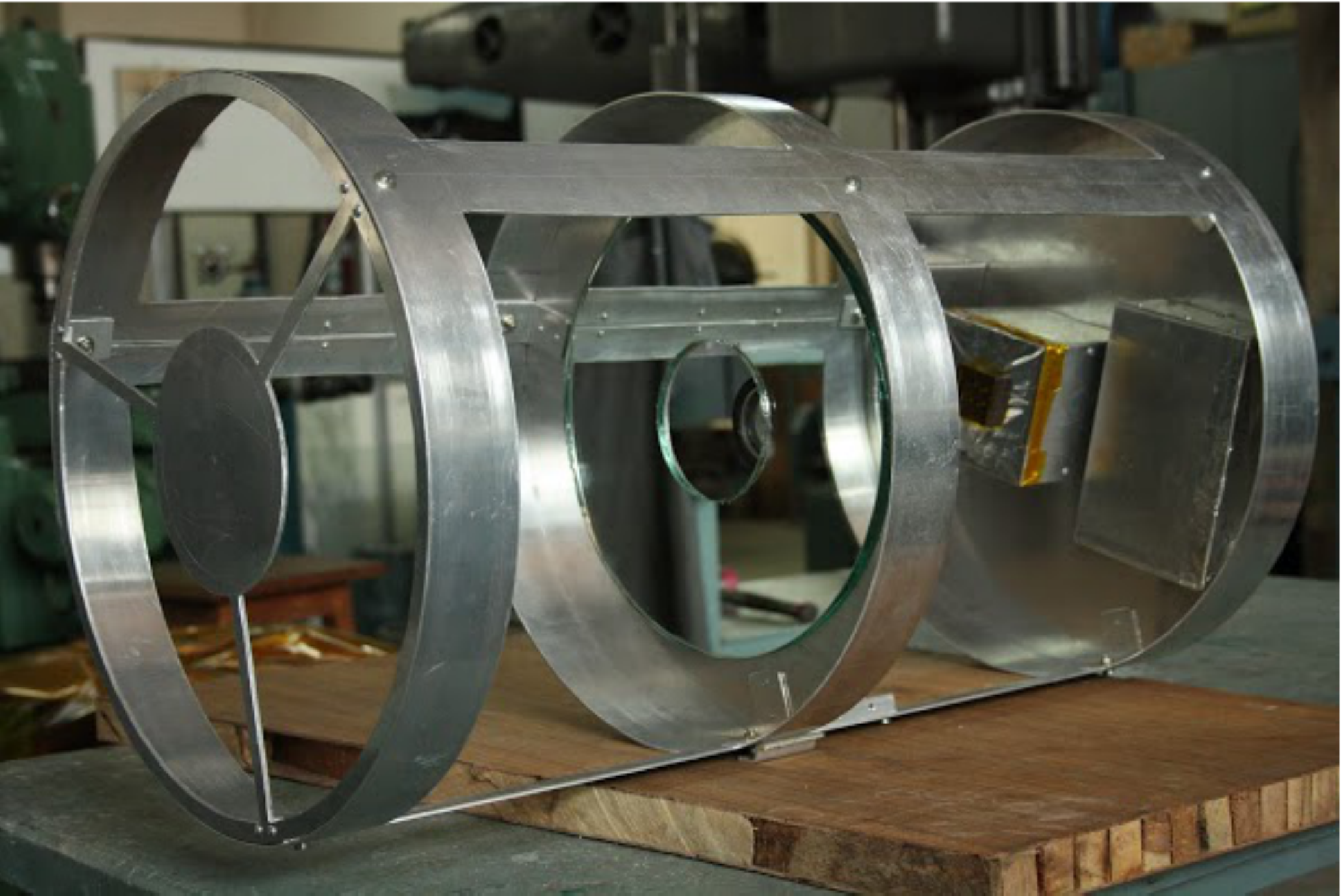}
\caption{{\it Left}: The optical layout of LUCI telescope performed   
using the optical design software {\it Zemax}. {\it Middle}: The mechanical and structural 
design of the telescope performed with {\it Solidworks} (mechanical design software). {\it Right}:
A photograph of the weight and volume model of LUCI.}
\label{fig:opt_design}
\end{figure}

The LUCI instrument is a 30-cm Ritchey-Chr\'{e}tien-based (Fig.~\ref{fig:opt_design})
telescope with a photon-counting detector. Ritchey-Chr\'{e}tien telescopes are 
well-suited for wide-field imaging in a compact package; the overall dimensions of the telescope
are $\sim 40\times 65$ cm. With the use of a carbon-fiber structures, a 30-cm primary mirror will weigh 
around 1.5 kg. We have designed for a $3$ degrees field of view with a pixel 
size of $16\mu\text{m} \times 22\mu$m, such that we can reach a limiting magnitude of 
21 magnitude (STMAG) in 1000 sec.

\subsection{Detector}

\begin{figure}[hb!]
\includegraphics[scale=0.2]{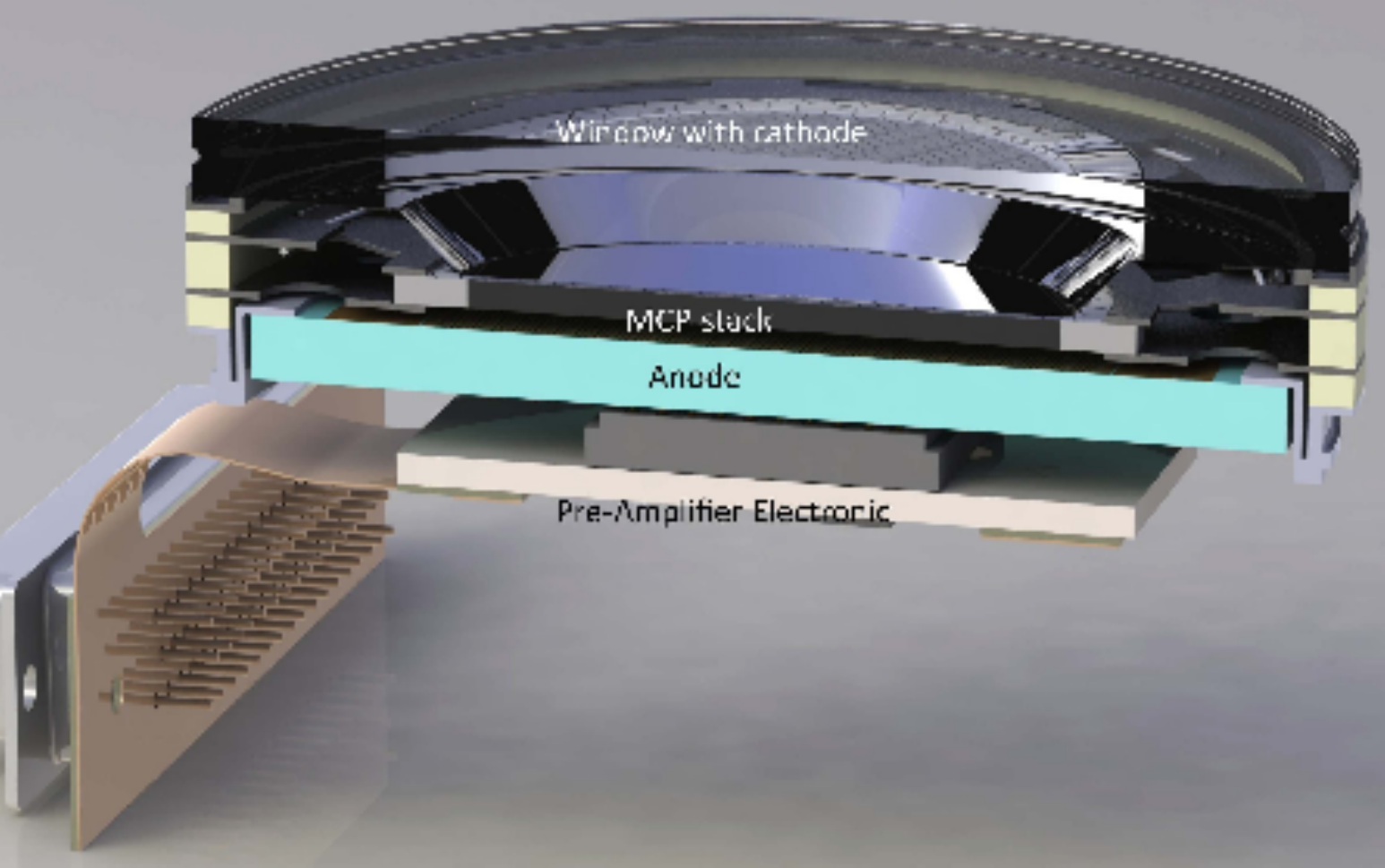}
\includegraphics[scale=0.5]{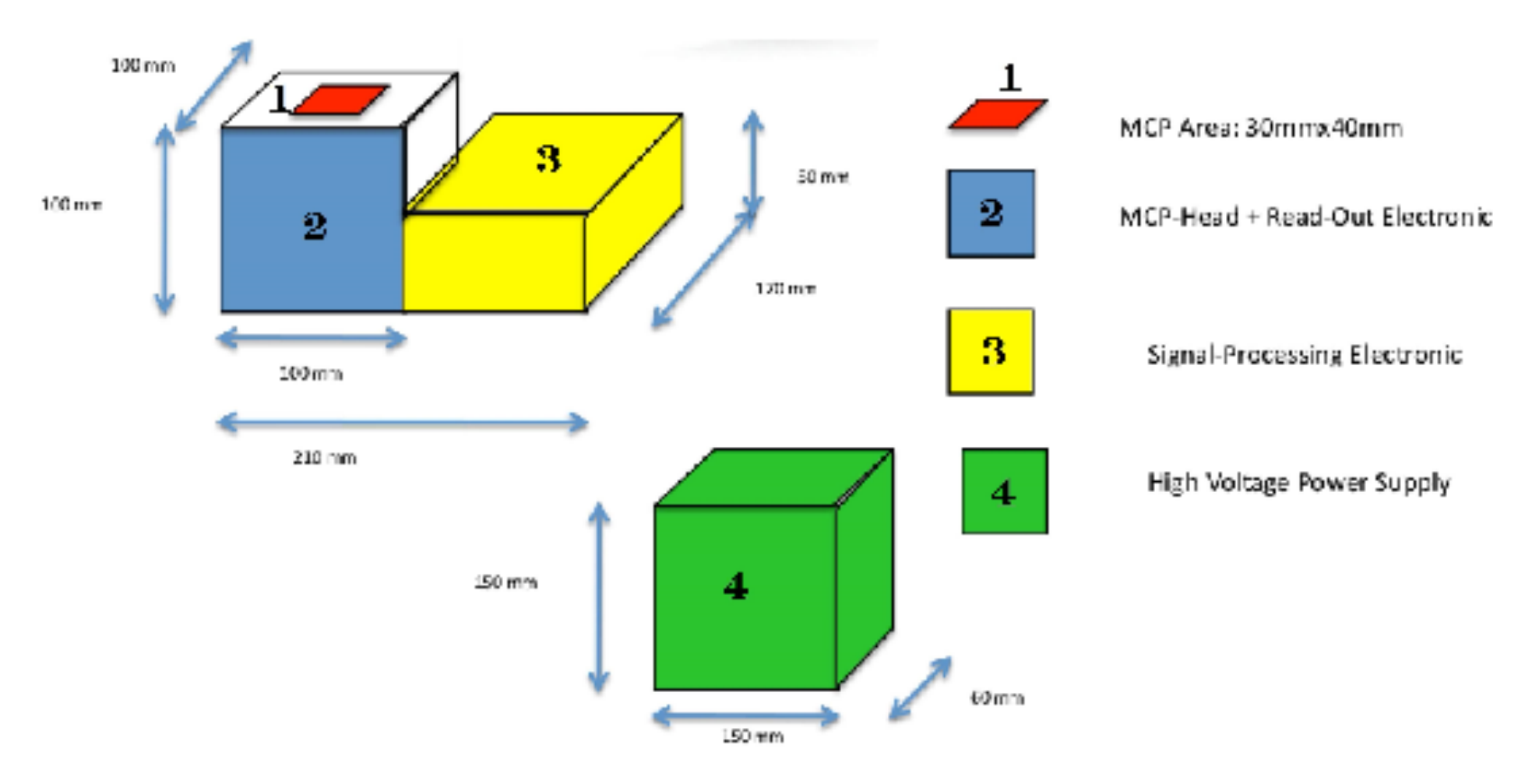}
\caption{Cross-strip anode based detector. {\it Left}: A detector head with the preamplifier
section but without the ``evaluation electronics" and high voltage power supply (HVPS). {\it Right}: 
Overall layout of the electronics module with dimensions. 1 --- MCP area. Box 2  --- the MCP head with the 
read-out electronics. The Signal-­Processing Electronic module (SPEM, box 3) can be placed 
beside or below the MCP-­Head. Position of the HVPS (box 4) as per requirements of the 
lander is below the SPEM. For the colour version of this figure please see the online publication.}
\label{fig:detector}
\end{figure}

The detector is a microchannel plate (MCP)-based single-photon counting detector 
(Fig.~\ref{fig:detector}). It consists of a caesium iodide (CsI) (optional gallium nitride GaN)
photocathode, a chevron 
stack of 2 MCPs and a position sensitive cross-strip (XS) anode sealed in a detector body.  
MCPs are two-dimensional periodic arrays consisting of many small-diameter glass channels; the 
channels of 2 MCPs are tilted with respect to the surface normal to suppress ion feedback 
from ionized residual gas. 
When a photon hits the cathode, an electron is released and is accelerated towards the 
MCP stack across a voltage gap of about 200 V. Each photoelectron produces secondary electrons 
by striking the channel walls of the MCPs resulting in an avalanche of electrons leaving 
the MCP stack at the bottom. A unique feature of an MCP electron amplifier is to provide 
high gains in excess of $\sim 10^5$--$10^6$ electrons per event while preserving 
the location within a single pore. Detectors requiring single event-counting capabilities 
utilize readout methods in which the electron cloud from the MCP is sampled by the 
anode and the charge information is processed by the front-end electronics (FEE) to 
determine the position where the original photon hit the detector and the photon arrival time.

The XS anode technique utilizes two orthogonal direction planes of parallel strips/wires 
that serve to detect $X$ and $Y$ positions. The XS anode works by dividing the charge rather 
than using a signal propagation time-delay, which give the capability of higher spatial 
resolution and a faster event-processing speed. The two axes layers ($X$ and $Y$ strips) 
collect an equal amount of charge from the MCP. The electron cloud from the MCP is collected 
on several strips (approximately five) on each axis which allows an accurate determination of 
the centroid of the event. Each strip of the anode is connected to an amplifier and then 
to an analog-to-digital converter (ADC), outside the vacuum, to read out the charge on 
individual strips. An advantage of an XS anode technique is good spatial resolution 
($<10 \mu$m FWHM) with a smaller MCP signal; it only requires two MCPs to achieve efficient 
gain ($10^5-10^6$ electrons). The detector also has a very fast signal propagation of $\sim 1$ ns. 

A chevron stack of MCPs with 40-mm active diameter, 80:1 L/D, $10^{\deg}$-biased microchannels 
with $10\mu$m pores on $12 \mu$m centers, is used in LUCI detector head. The XS-anode 
has an active area of $33\times 44$ mm$^2$ with $64\times 64$ strips. Each active set has 64 
strips separated by an insulating layer. The strips of the two active sets are perpendicularly 
aligned. Due to the insulating layer and the spacing between the strips, the 
sensitive area is around 62\% of the total area. These two orthogonal strips share (roughly 
equally) the charge generated by the MCP stack placed $\sim 2$ mm above it. The analog signal 
streams provided by a special pre-amplifier chip (which receives the anode signals), are 
digitized by four analog-to-digital converters, which are part of the FEE. Digital event 
information is processed in a field-programmable gate array, which calculates the location 
where the original photon hit the detector, the total charge of the event, and a time stamp 
for the event (maximum 300,000 photons/second). 

The weight of the entire detector system (detector head, electronics and HVPS) is around 3 kg 
and the power consumption is less than 10 watts. The overview 
of the system and the projected performance is summarized in Table~\ref{table:overview}.

This detector is developed by the Institute of Astronomy and Astrophysics of the University of 
T\"{u}bingen, Germany (Kappelman et al. 2006). The development has the heritage of two
successful ORFEUS-SPAS 
``free-flyer" missions (Diebold et. al. 2012) and will be space-qualified onboard the next German 
TET-satellite (technology test-bed satellite). In the advanced design phase we will study the necessity 
of processing the event stream into images, so as to reduce the required downlink 
telemetry volume.

\subsection{Delivery and platform}
\label{sec:shielding}

\begin{figure}[hb!]
\includegraphics[scale=0.3]{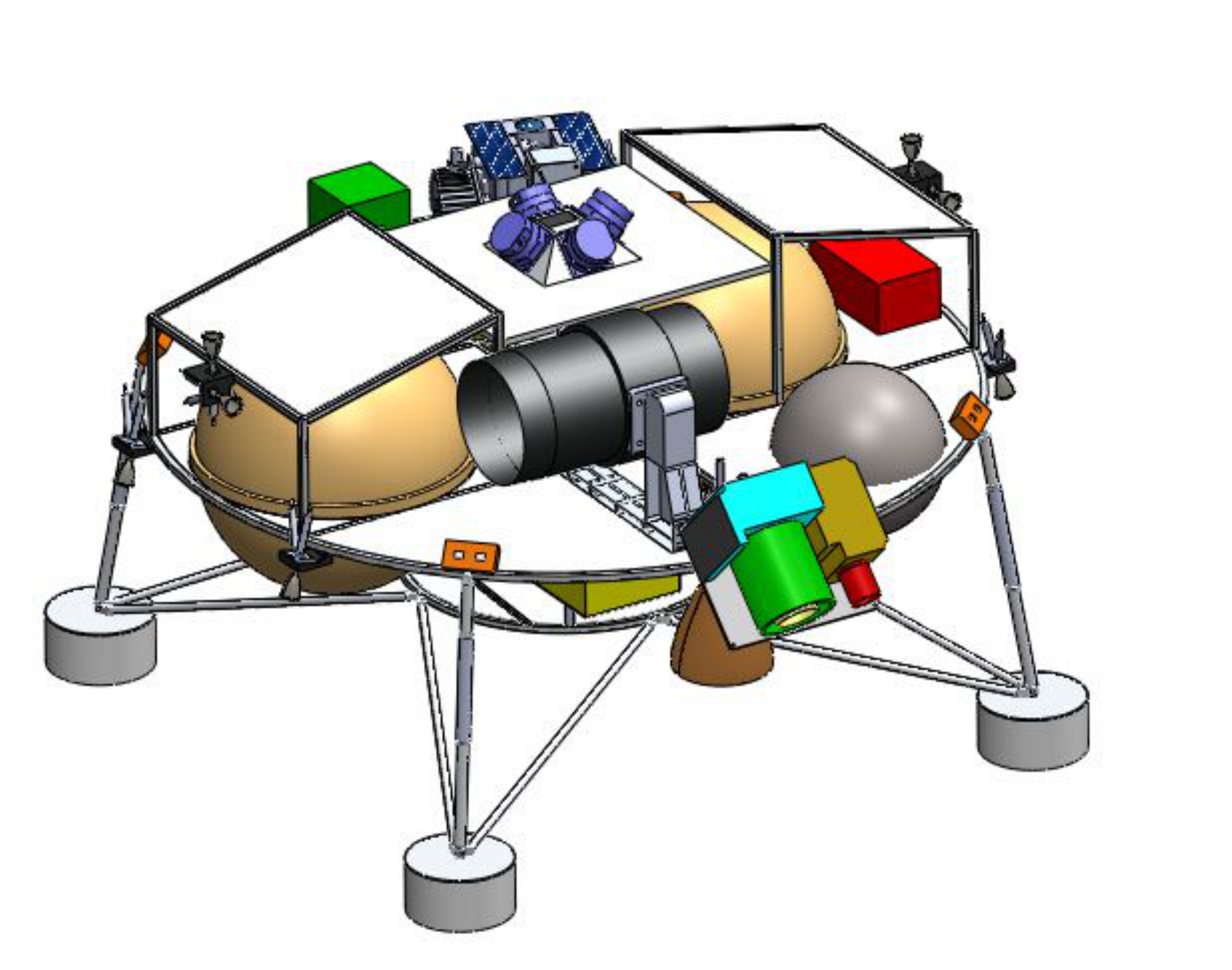}
\includegraphics[scale=0.25]{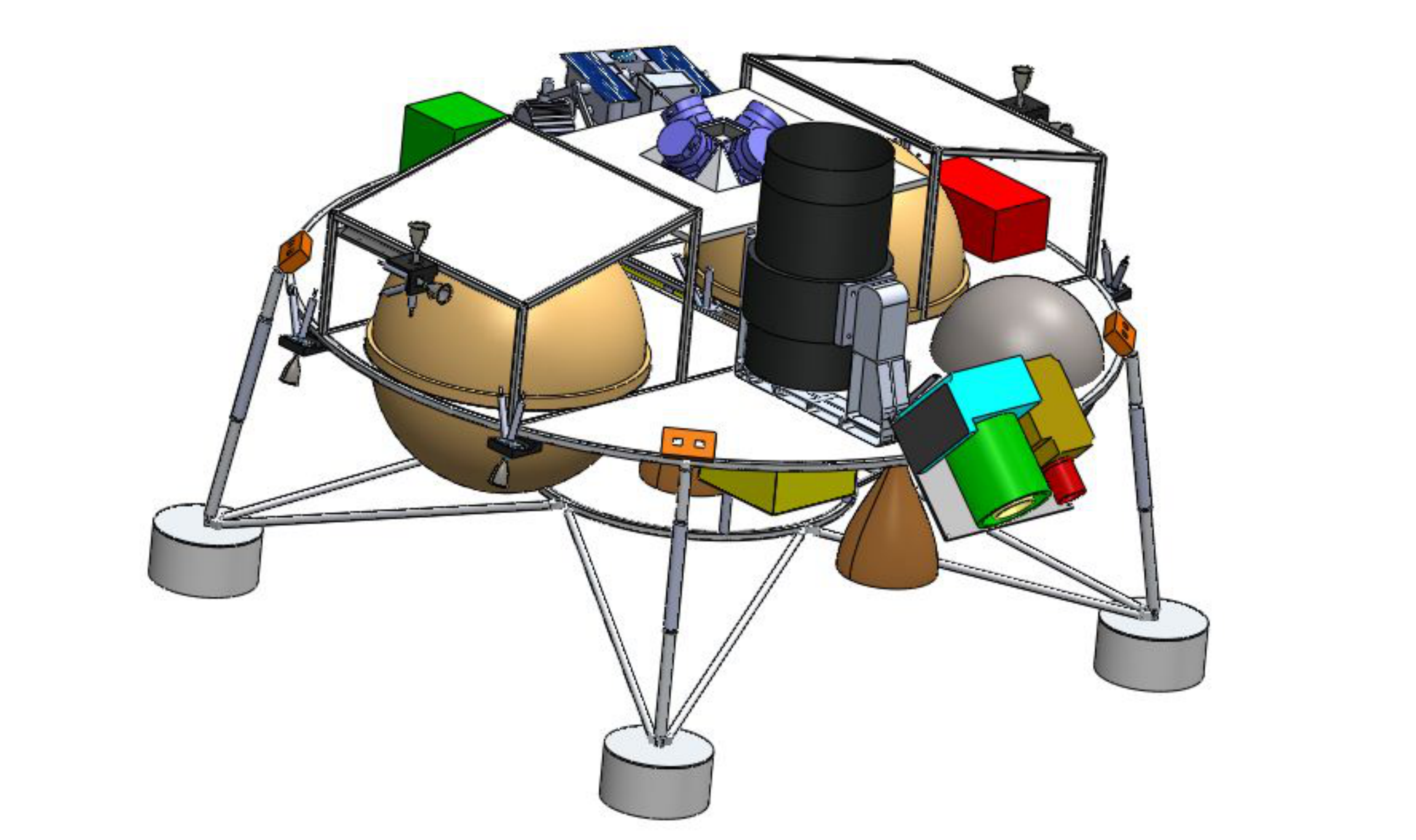}
\caption{{\it Left}: The preliminary design of the lander and placement and deployment of LUCI 
(black cylinder); {\it Left}: stowed position, {\it Right} default operation position. Courtesy of the 
Team Indus.}
\label{fig:lander}
\end{figure}

The total weight of the complete instrument has to be 7--8 kg due to the 
mass restrictions imposed by the carrier. The telescope will be mounted 
on the upper deck of the lander (Fig.~\ref{fig:lander}) and may have the pitch angle 
freedom of rotation, with zenith angle range of $+30^{\circ}$ to $-30^{\circ}$ due the
constraints explained in Sec.~\ref{sec:ObsMode}.

\paragraph{Power and telemetry}
\label{sec:power}
 
The required power for the detector will be provided by the platform where solar 
panels are the primary source of electrical energy during lunar day surface operations,
therefore LUCI will only operate in the daytime. The HVPS will be used for the required high 
voltage for the detector and the input to the the power supply will be a standard logic voltage. 
The operational power requirement for the detector is less than 10 watt. 

The platform will provide a shared downlink data telemetry in addition to a limited command uplink
facility; the total available datalink budget on the platform is 200 Kbps downlink and 100 bps uplink. 
We will use the uplink to readjust the observational plans in case the lander misses the programmed 
location and position.
 
\paragraph{Storage and deployment of the telescope}
 
The telescope will remain in a stowed horizontal position for the entire duration of the flight from launch to
touchdown, protected from mechanical shocks, radiation and sunlight amongst other environmental conditions
during transit by the housing provided by the platform (Fig.~\ref{fig:lander}, {\it Left}). 

The telescope's operations will begin after the touchdown and release of the rover. The initial deployment will be
at a pitch angle of $+90^{\deg}$ to assist in the attitude verification of the lander 
(Fig.~\ref{fig:lander}, {\it Right}).  

Since LUCI is planned to be mostly contained within the lander, we will need only 
minimal thermal blankets, since the platform will take care of most of the insulation, and the situation 
is similar for the radiation shielding. The aperture will be well baffled to avoid reflections and scattering 
from the outer surfaces of the lander. In order to ensure the survival over the lunar night, the telescope 
will be lowered back into its storage bay and the detector will be switched off. The platform will have a 
minimal battery power to provide the heater.  

\paragraph{Dust} 

There has been considerable interest in lunar dust --- produced from micrometeorite impacts and
electrostatically charged by solar photons --- with the LADEE spacecraft 
designed to study its characteristics. Dust raised by the lander may rise to
considerable heights and settle back down on the LUCI mirrors. We expect only a limited
impact from this dust, however, as the motion of the dust is largely across the terminator 
(Berg et al. 1974) with illuminated surfaces likely to be covered and cleaned by the same process. 
In fact, the Lunar Laser Ranging (LRR) reflectors left by the Apollo astronauts are still 
working after more than 40 yrs albeit with $25\%$ efficiency (Lester 2007). They were 
neither shielded from direct sunlight nor from dust. The baseline LUCI's operation is no more than a 
few months and we do not expect dust accumulation on the mirrors to be a problem. Moreover, the 
dust levitated by the landing itself is not likely to settle on the LUCI optics since at landing the telescope 
will be in the stowed position in the storage bay.

\subsection{Bandwidth, resolution, sensitivity}

The LUCI instrument requires high-sensitivity imaging in the FUV but requires the rejection of
longer wavelengths. In addition, we require the blocking of wavelengths shortward of 127 nm 
to cut off the Ly-$\alpha$ (121.6 nm) line. We are not using filters in LUCI, thus the cut-offs are provided 
by the detector and the optics. To meet the demands for the bandpass, the detector has 
a magnesium fluoride (MgF$_2$) window and a CsI photocathode. CsI 
provides good responsivity in all UV region and is solar blind with 180 nm cut-off on 
the long side. We plan using CaF$2$ corrector lenses which will block light below 135 nm.

A typical sampling of $\sim 5.3^{\as}$/pixel with a total field of view of 3 degrees yields the 
spatial resolution of about $16^{\as}$.

\subsection{Calibration}

\subsubsection{Ground calibration}

The ground calibration of the LUCI instrument will be performed at the M.~G.~K. Menon
Space Science Centre at the CREST campus of the Indian Institute of Astrophysics, Bangalore, India.
This facility is now being used for the integration, characterization and calibration 
of the UVIT instrument (Kumar et al. 2012). The telescope assembly and alignment will be 
performed in class 1000 clean rooms with laminar flow tables providing a class 100 or 
better local environment. Procedures will be adapted from the UVIT calibration and, 
where possible, equipment from the UVIT program will be used.

\subsubsection{Astrometric calibration}

The astrometric calibration of LUCI will begin with the absolute localization of the lander, which
is described by its location in selenographic coordinates $\lambda$ and $\phi$, latitude and longitude, respectively,
and the position, determined by the 3-axis attitude: pitch, roll and yaw angles. The yaw angle is defined 
as the LUCI pointing. The lander will be equipped with inclinometers to determine its pitch and roll angles. 
In addition to six Sun sensors that will provide the location of the Sun, we will use LUCI to determine the
offset of the lander's zenith from the `true' zenith by beginning observations at the pitch angle of $90^{\deg}$. 
The lander will have a precise on-board time clock and Moon orbital information, which together with the positions of the
Sun and the Earth (determined by the Earth antenna) will provide the value of the yaw angle, i.e. the LUCI
azimuthal pointing. The programmed LUCI pointing is $\text{Az} = 130^{\deg}$ at programmed landing site of $\lambda=+38.336^{\deg}$, $\phi=26.006^{\deg}$W. 
In case of missing the programmed localization, we will update the observational plan by uplink from the Earth.

\subsubsection{Photometric calibration}

The primary photometric calibration of LUCI will be made with reference to a set of 
well-studied hot, young and/or massive stars, including bright $O$, $B$ and hot white 
dwarfs (WDs) as was done for the UV instruments onboard the HST. These stars are 
photometrically stable, their fluxes are well matched to the sensitivities of modern 
space instrumentation and their spectral energy distributions (SED) are well 
characterized. The FUV fluxes of these stars are presently known to better than $10\%$. 
The standards will be observed as they transit across the LUCI FOV.

For the definition of LUCI photometric system, we will use the AB magnitude 
system that is widely used in UV astronomy (Oke \& Gunn 1983)
\be
m_{\rm LUCI}=-2.5\log{\bar{F}_{\nu}} -48.6\,,
\ee
where $ \bar{F}_{\nu}$ is averaged monochromatic flux (in ergs/sec/cm$^2$/Hz) at the 
effective wavelength (averaged across the bandpass),
\be
\bar{F}_{\nu} =\fr{\int_{\lambda_1}^{\lambda_2} f_{\nu}(\lambda) R(\lambda) 
d\lambda}{\int_{\lambda_1}^{\lambda_2} R(\lambda) d\lambda}\,,
\ee
where $R(\lambda)$ is the total system's response, $f_{\nu}$ is the SED of a source, 
and $\lambda_1$ and $\lambda_2$ are the lower and upper wavelength limits of LUCI's 
bandpass. However, since LUCI is a photon counting device, the real units of LUCI are 
photon counts per sec, that can be converted accordingly to flux units or magnitudes 
using 
\be
\text{Counts/sec/\AA} = F(\text{ph/cm}^2/\text{sec/\AA}) \times A_{\rm eff} \,,
\ee
where $F$ is source flux and $A_{\rm eff}$ is the total effective area in cm$^2$ --- 
the product of the mirror geometric collecting area ($A_{\rm coll}$), detector quantum efficiency (Q.E.),
and the optics efficiency which includes reflectivity of mirrors, 
absorption by the correcting lenses and absorption by the metering structures (`spiders') 
holding the secondary mirrors,
\be
A_{\rm eff} =  A_{\rm coll}\times \text{Detector Q.E.} \times \text{Optics Efficiency}\,.
\ee
The effective area $A_{\rm eff}$ of LUCI will be determined through the ground calibration tests.

\begin{table}[ht!]
\begin{center}
\caption{LUCI Instrument Overview and Projected Performance}
\vskip 0.1in
\label{table:overview}       
\begin{tabular}{l|l}
\hline\noalign{\smallskip}
\multicolumn{2}{c}{\bf LUCI Instrument Overview}       \\
\noalign{\smallskip}\hline\noalign{\smallskip}
\multicolumn{2}{c}{Optical module (OM)}			   \\
\noalign{\smallskip}\hline\noalign{\smallskip}
Aperture                  & $30$ cm   		   \\
Optical design            & Ritchey-Chr\`{e}tien   \\
Focal length   		  &$ 945$ mm,  F\#3.15  \\
FOV           		  &    $3^{\circ}$   	   \\
Mirrors coating	 	  & AlMgF$_2$  		   \\
Geometric collecting area &$\sim 607$ cm$^2$ 	   \\
Passband       		  & 130--180 nm	   \\    
Primary mirror   	  & 30 cm dia, mass $1.3$ kg 		   \\
Secondary mirror  	  & 9 cm dia, mass $\sim 0.5$ kg 	   \\
Supporting structure  & mass $\sim 2$ kg 	   \\
\hline						   \\
Total weight & $\sim 4$ kg 	   \\
\hline\noalign{\smallskip}
\multicolumn{2}{c}{Detector module }			   \\
\noalign{\smallskip}\hline\noalign{\smallskip}
Detector          	  & Sealed-tube 2-stack MCP with XS anodes\\
Detector window           & MgF$_2$  \\
Detector photocathode     & CsI opaque (Option: GaN semi-transparent)  \\ 
Detector size             & $33$ mm$\times 44$ mm   \\ 
Pixel size                  & $16\mu\text{m} \times 22\mu$m  \\
Total size in pixels      & $2048\times 2048$        \\ 
Detector QE               & CsI:  5--30\% (GaN: tbd)	    \\
Detector dynamic range   & 300,000 cts/s            \\
Detector power 		        &  $< 10$ W		    \\
Dimensions (cm)              & $21\times 17\times 10$  \\
\hline						   \\
Total weight (incl. FEE and HVPS)           & $\leq 3.5$ kg		   \\
\hline\noalign{\smallskip}
\multicolumn{2}{c}{Projected performance}	   \\
\noalign{\smallskip}\hline\noalign{\smallskip}
Resolution                               & 	$\sim 16^{\as}$  		                            \\
Pixel scale                               &  $5.3^{\as}$/px	                                  \\
Time resolution (nominal)          & few ns               	                                  \\
Sensitivity (point source)            & $21^m$ monochromatic in 5,000 sec            \\
Exposure time$^{\dag}$           &  $10,800\times\sin{(\text{Alt.})}$\\
\hline						                                                                         \\
Total LUCI weight                            & $\sim$ 7--8 kg 	                                         \\
\noalign{\smallskip}\hline
\end{tabular}
\end{center}
$^{\dag}$ {\small The exposure time on a single object can be expressed as $10,800 \sin{\psi}$ sec, 
where $\psi$ is the altitude (this is the time it takes a point source to cross the centre of a $3^{\deg}$ FOV).
The total area covered during one lunar day is $178^{\circ} \sin{\psi}$,
which gives $102^{\circ}\times 3^{\circ}$ at $35^{\circ}$ from horizon.}
\end{table}

\section{Scientific objectives}

We have identified several science areas where the imaging far-UV observations from the 
Moon can deliver unique scientific results. However, it should be noted that, in many ways,  
this is a pioneering experiment, part of a new way of doing astronomy. The cost
of going to space is decreasing rapidly and there may be many opportunities for serendipitous science;
perhaps as suitcases on flights of Virgin Galactic\footnote{Virgin Galactic is a company within Richard
Branson's Virgin Group which plans to provide commercial space flights.} or as nano-satellites 
based on the cubesat format (Brosch, Balabanov \& Behar 2014). This particular opportunity
takes advantage of the Team Indus effort to send a rover to the Moon. We have carefully selected a
set of science objectives and instrumentation suited for this first opportunity. However, as more and
more such flights occur, we expect that astronomers (and other scientists) will be able to take
advantage of a diverse set of missions to perform unique science.

\subsection{Advantages of FUV}

The FUV sky is relatively dark and is dominated by a small number of discrete UV sources
such as hot stars, particularly {\it O} and {\it B} stars, hot WDs and subdwarfs, AGN and 
star-forming galaxies. 

A very important advantage of a lunar FUV telescope with a wavelength range limited
to 135--180 nm and the use of a CsI cathode is the absence of several noise contributors 
which are intrinsic to telescopes at longer spectral regions in LEO:
\begin{enumerate}
\item Zodiacal light ---  sunlight scattered by interplanetary dust contributes significantly at 
wavelengths longward of 200 nm, where the solar flux rises rapidly and practically negligible
in the FUV range.
\item  Airglow --- line emission from the Earth's upper atmosphere bright throughout
the UV range (see Fig.~\ref{fig:appollo}) and is an important background contributor 
for LEO missions (Murthy 2014).
\item Instrument background --- the dark count in UV instruments due to fast
particles hitting the MCP. Though typically low, it varies throughout the orbit 
for LEO spacecrafts, rising in areas of high particle count, such as the South Atlantic 
Anomaly or the polar regions. On the Moon, Galactic cosmic rays and solar wind particles 
are the main sources of radiation. Their flux, however, is about half of that in 
deep space due to self-blocking by the Moon (Adams et al. 2007). 
\item Spacecraft jitter --- a lunar platform is very stable, thus one more advantage in 
comparison with the space-based telescope.

\end{enumerate}

\subsection{Astronomy in the far-UV domain}

When discussing astronomy in the FUV domain, one has to consider first what are the sources 
expected to emit significant amount of radiation in this spectral region. In general, sources 
will produce either continuum radiation or discrete emission lines. The continuum can either be 
from a black-body (BB) or almost a black-body (such as a star or a population of stars), or from 
a non-thermal source (NTS), such as synchrotron emission from electrons in a magnetic field.

A BB is characterized by its effective temperature. In order to produce a significant flux 
in the LUCI bandpass, a stellar source has to be 3--4 times hotter than the Sun, 
18000 to 24000$^{\circ}$K, corresponding to early-$B$ or $O$-type stars. Note also that 
late-$B$ and even $A$-type stars or hotter may be detected, 
provided they are relatively nearby. Stellar radiation may be ``blue-shifted'' and 
become more UV-dominated if it is scattered by small dust particles.
This is the inverse of interstellar reddening and can be seen 
in the vicinity of hot stars that are associated with dust as reflection nebulae. 
A good example of a reflection nebula is NGC 1435 near 23 Tau, the Merope star in the 
Pleiades, with its brightest knot IC 349 (Herbig 1996; Herbig \& Simon 2001).

Just like single stars, entire galaxies can significantly contribute to the FUV continuum radiation, 
provided their stellar population contains large numbers of $O$ and early-$B$ stars. Since 
these stars are massive, 10 M$_{\odot}$ or more, they are also short-lived. The FUV is, 
therefore, a good indicator of very recent star formation. By ``counting'' the FUV photons one 
can, therefore, account for the massive and short-lived stars in an extended source,
revealing a very recent star formation burst.

Extended extragalactic FUV sources are often produced from the interaction of 
galaxies with the intra-cluster 
medium in clusters of galaxies. Examples of such phenomena are in the Virgo Cluster
(NGC 4330; Abramson 
et al. 2011), and in the Coma cluster (Smith et al. 2010). However, 
in some cases extended emission is produced by star formation away from the 
main body of a galaxy. Examples 
for this are the UV filaments detected near brightest cluster galaxies (Oonk et al. 2011) 
and the extended UV disks of some galaxies (e.g. Thilker et al. 2007).

In the Solar System, the expected FUV sources are the Lyman $\alpha$ halos around planets produced by 
resonant scattering of the solar radiation, which however are outside the proposed LUCI band, 
and aurorae and airglow in the atmospheres of planets and major satellites. A few FUV observations of
comets were made by GALEX (e.g. Morgenthaler et al. 2011), where the emission 
is dominated by the CI emission bands excited by the solar radiation and wind.

\subsection{Lunar sky and modes of observation}
\label{sec:ObsMode}

As shown in Sec.~3, a stationary telescope on the Moon scans the sky and over a
period of one lunar day approximately 44 deg$^2$ of the sky can be observed. 
The baseline operation is with the zenithal orientation of LUCI (the ecliptic is always at 
$\sim 40^{\circ}$ from zenith), however, we will be able to change the pitch angle between 
$0$ (horizon) and $90$ (zenith) degrees. 

Observations will be performed using the altitude/azimuth (Alt/Az) coordinate system. 
The celestial coordinates of the object will be converted to the attitude of 
the telescope pointing, i.e. the azimuth and the altitude angle of an object on the lunar sky. 
At low altitudes, star motions are no longer tangential and thus the tracks have a 
small curvature. To avoid this problem we will be pointing the telescope at a typical 
altitude angle of more than $30^{\circ}$. Pointing above this angle also removes the 
problem of the lunar ``horizon glow" - a reported scattering of sunlight 
by lunar dust ``atmosphere" (Murphy \& Vondrak 1993).

\begin{figure}[ht!]
\includegraphics[scale=1.1]{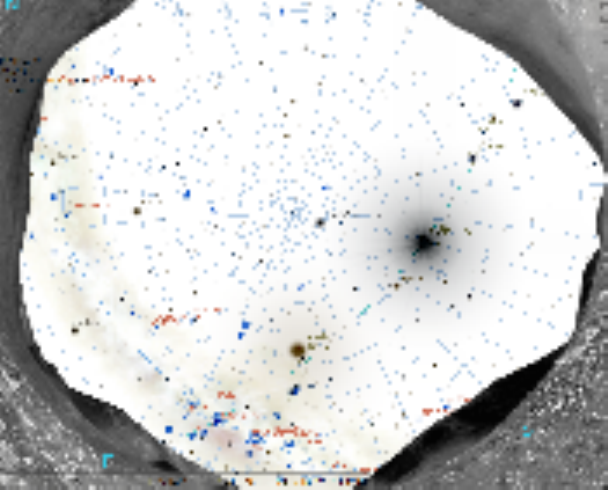}
\includegraphics[scale=1.]{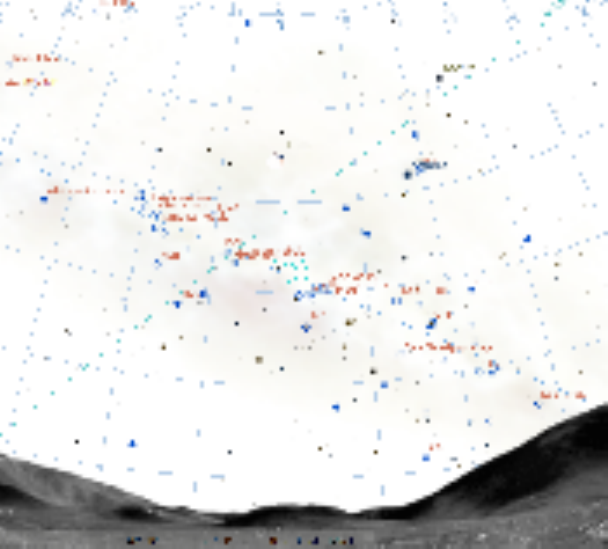}
\includegraphics[scale=1.1]{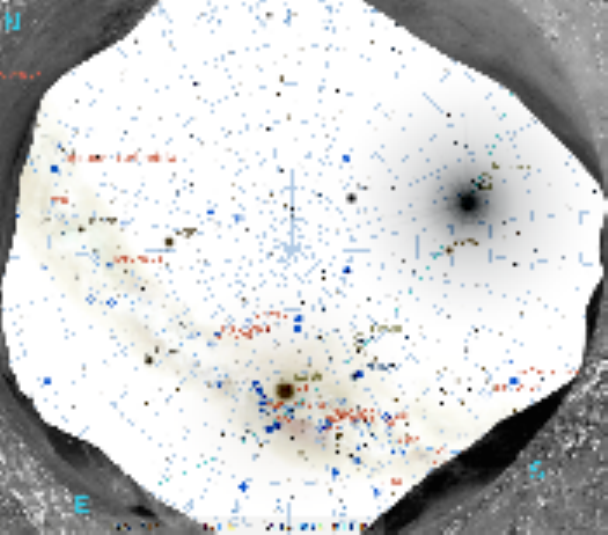}
\includegraphics[scale=1.1]{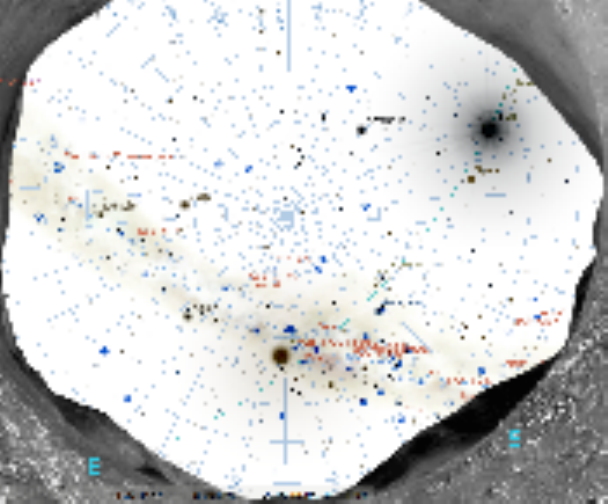}
\caption{View of objects in the lunar sky on first few days of LUCI's operations with simulated
lunar terrain. The grid is the horizontal coordinate system (Az/Alt), with zenith in the centre. {\it Top left}: 
02 Oct. 2015, 03 hrs; {\it Top right}: 03 Oct. 2015 13:00 hrs, map is centred on the Galactic centre
(blue circle at $\sim 32^{\circ}$ elevation; Earth is above at $\sim 43^{\circ}$) 
crossing the $130^{\circ}$ azimuth line. {\it Bottom left}: 04 Oct. 2015 noon; {\it Bottom right}: 
05 Oct. 2015, 03 hrs. The green line is the ecliptic and the blue line is the Galactic plane (see online 
publication for the colour version of this figure).}
\label{fig:maps}
\end{figure}

The proposed landing site is at Mare Imbrium/Sinus Iridum at selenographic coordinates 
$38.336N$ and $26.006W$. The accuracy of the landing is about 100 km. 
We have generated maps of the lunar sky for the months of October 2015 -- 
December 2015 using the {\it Stellarium} software. In case the lander can 
orient before the actual landing to a pre-determined direction, we select the azimuthal angle of  
$+130^{\circ}$. At this orientation,
the Earth crosses the $130^{\circ}$ azimuth line (see Fig.~\ref{fig:maps}, {\it Top right}) 
and we will be able to observe it. Earth is 
bright in the FUV, in particular, the geocorona extends to $\sim 80,000$ km emitting 
large amount of Lyman-$\alpha$ photons at 121.7 nm. It would be very interesting part to 
observe the Earth at different phases at small illumination (in the lunar day the Earth is
never in the full phase) in the UV range outside the Lyman-α contribution.
Since the worst case scenario is that LUCI would not survive the first lunar night, we plan to 
observe important targets in the first month: calibration objects, Galactic plane, 
Earth --- we provide detailed maps of the first months as online source and maps for the first few days 
in Figure~\ref{fig:maps}. For example, on 02/10/2015 Earth is at Az/Alt: $\sim 129^{\circ}/44^{\circ}$
at $20\%$ illumination. On 03/10/2015 at 13:00 hrs, the Galactic Centre is at Az/Alt:
$\sim 130^{\circ}/32^{\circ}$ and Earth is at Az/Alt:$128^{\circ}/44^{\circ}$ at $34\%$ illumination. 
On 04/10/2015 at 02:00 hrs, the intersection of the Galactic plane with the ecliptic crossing the 
$130^{\circ}$ azimuth at $39^{\circ}$ elevation. All times are India Standard Time (IST), which is
UTC $+5.5$ hours.

\begin{figure}[h!]
\includegraphics[scale=0.35]{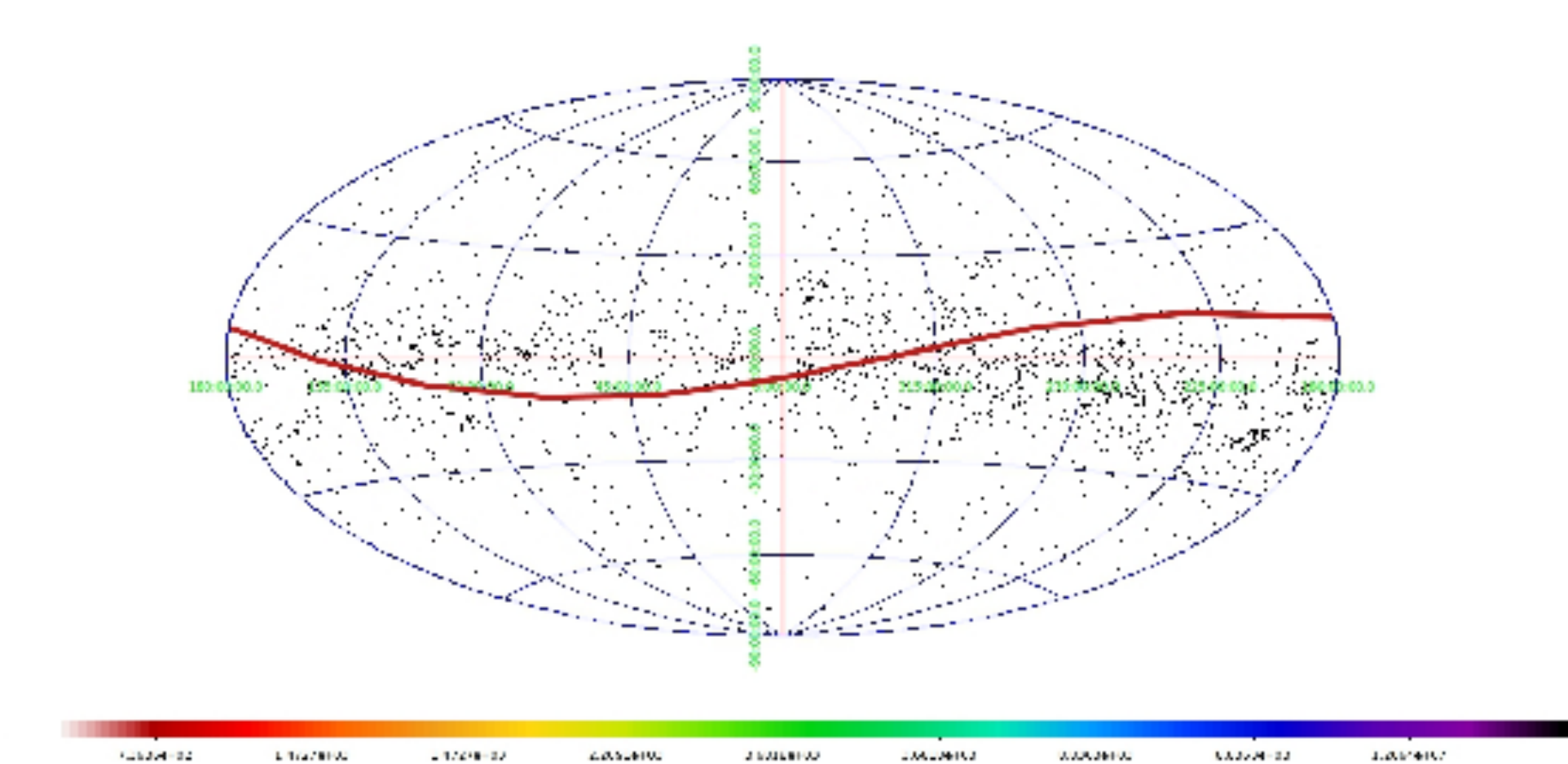}
\includegraphics[scale=0.35]{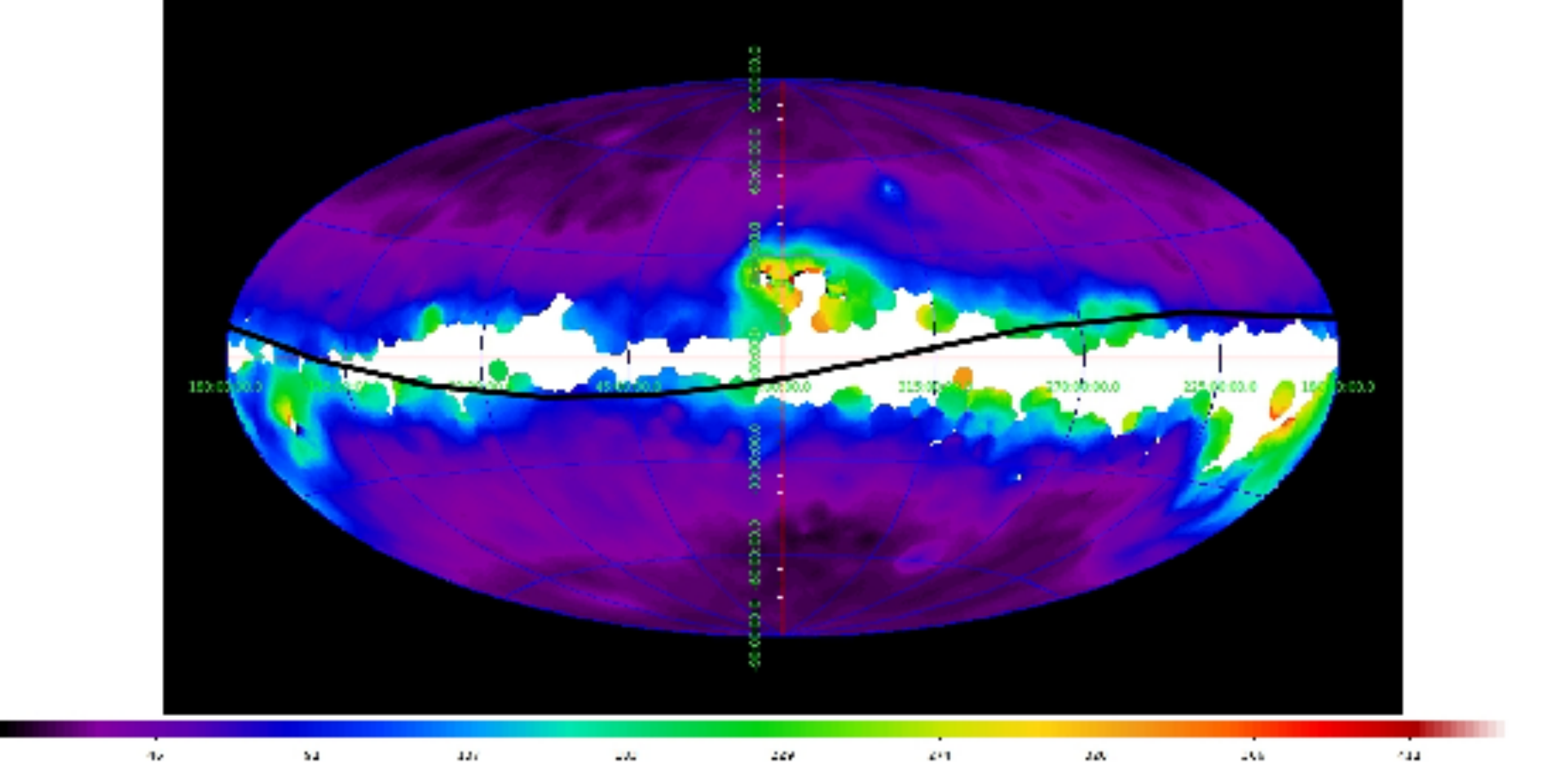}
\caption{Simulations of the sky background for October 2015 in FUV in a projected 
bandwidth of LUCI: stars and diffuse Galactic light (top and bottom panels, respectively). 
Since no GALEX data is available for the regions in and around the Galactic plane, we use a 
cosecant law to simulate the diffuse background in these regions. The solid line is the local horizon 
--- only the sky above this line is visible to LUCI in October 2015. The following ASTUS inputs 
were used: dark current=0 cps, field of view$=3^{\circ}$, AITOFF projection, pixel size = 
$0.005^{\circ}$.}
\label{fig:astus_simulations}
\end{figure}

The importance of properly accounting, if not of eliminating, the diffuse backgrounds when 
dealing with extragalactic observations was stressed by Mattila (2006). Scattered light 
due to dust clouds on the Moon can play the role of a sky background but the choice of 
the FUV range removes this problem. We have simulated the far-UV background
with LUCI's parameters using the UV sky background simulator ASTUS\footnote{available at 
{\tt http://cads.iiap.res.in/tools/uvskyCalc}} (Safonova et al. 2012). In 
Figure~\ref{fig:astus_simulations}, we display the results of all-sky simulation for the 
month of October 2015 in a projected bandwidth of LUCI assuming a flat step filter profile 
for two separate components of the background: stars and diffuse Galactic light 
(top and bottom panels, respectively). The thick solid line marks the local horizon; only the 
sky above the line is observable in October. For example, we can observe the Galactic plane, but 
not the Magellanic Clouds. We have also calculated the count levels 
(diffuse Galactic + stars) for a few open clusters that can be used for astrometric 
calibration. Astrometric calibration includes pointing accuracy estimation, FOV calibration, 
distortion correction, etc. For these purposes, extended fields containing many stars with 
good astrometry, such as open clusters, are generally used. Zodiacal light is usually nearly 
negligible in FUV, therefore we did not consider it in these simulations. The background
levels produced by diffuse light and stars in the fields of possible calibration targets in 
counts/s over the full FOV are given in Table~\ref{table:astrometry}. ASTUS uses Hipparcos 
catalog for modelling of stars, which however, does not contain white dwarfs. 

\begin{table}[hb!]
\begin{center}
\caption{Diffuse background (b/g) and stellar light levels in counts/s in the fields of
possible photometric and astrometric calibration targets for LUCI.}
\vskip 0.1in
\label{table:astrometry}
\begin{tabular}{llclll}
\hline
{\bf Phot. Standard}&m$_{\rm V}$&{\bf Sp.~type}  &{\bf Diff. b/g}&{\bf Stars}\\\hline
HD 60753            & 6.7        &  B3IV     & 15262.9  &1859.2     \\\hline
HD 49798            & 8.3        &   O6       & 12854.1 &2734.6      \\\hline
HR 1996             & 5.17       &  O9V      & 2005.3   &11777.1      \\\hline
{\bf Open Cluster}  &{\bf Dim}$_{\rm V}^{\dag}$ &{\bf No. of stars} & & &       \\ \hline
Feigelson1          &     $18$  &10           &5963.87   & 37828.5   \\ \hline
ASCC 51             &     $48$  &18          &7902.4    & 45231.7   \\
\hline
\end{tabular}
\end{center}
$^{\dag}$\footnotesize{Apparent dimension in $V$ in arcminutes.}
\end{table}

\section{Primary science goals}

\subsection{Solar System bodies in UV}

\subsubsection{Earth in FUV}

The historical image of the Earth in the FUV (Fig.~\ref{fig:appollo}) made by the Apollo 16 
team 40 years ago is still interesting in terms of a possibility to study/monitor the 
Earth’s upper atmosphere that could benefit communications and help measure long-term 
effects of air pollution. It also shows the occultations of stars by the Earth's atmosphere --- another interesting 
possibility to consider. Earth occultations can be used just like lunar occulations, e.g., analysis of 
lunar occultations have discovered some stars to be very close visual or spectroscopic binaries.  In addition,
angular diameters of stars can be measured by timing of occultations, which is useful for determining 
effective temperatures of those stars. The key is the time-resolved photon detection that will allow
seeing stars entering the atmosphere and leaving it at the end of the event.

\subsubsection{Near-Earth Objects}

{\it ``The dinosaurs became extinct because they didn't have a space program."}, Larry Niven. 

At present, there are approximately 620,000 asteroids whose orbits are known in our Solar System; this
is less than $1\%$ of the estimated number of objects that orbit the Sun. And less than $10\%$ of all NEO have 
been discovered, according to the NASA scientist Donald Yeomans (2013). 

Till now most spectral reflectance studies of asteroids and comets were done in optical and NIR, however, recent lab
measurements, space missions and HST remote observations showed (Li et al. 2011) that the asteroidal
UV region contains a wealth of diagnostic information, which is not yet well understood. Some
asteroids have UV absorption bands and some do not. Asteroids are subject to space weathering,
which is believed to reducing their optical albedo, suppressing absorption features in optical and NIR
(Chapman 2004), but brightening them in the UV (Hendrix \& Vilas 2006). Yet only a few scattered
observations of some asteroids in the UV exist apart from some 50 IUE collected spectra in the 1980s
(Li et al. 2011b). In spite of the theoretical work on the UV reflectance of the meteoroids and
IUE UV observations, the understanding of asteroid UV reflectance is limited due to the insufficient 
space and lab data (see a report on measurement of the FUV reflectance of asteroid Lutetia by {\it Rosetta} in 
Stern et al. 2010).

Apart from being a menace, near-Earth objects are important in their own right, 
providing information on the original environment of pre-solar nebula. They may also have played a
key role in starting of life on Earth, since asteroids and comets are thought to bring to the Earth
much of the water and carbon-based materials. They are rich in minerals, metals and their clays 
and ices could provide water resources, thus the NEOs may serve as supply and fuel stations for 
future space flights. It is important for human survival to continue to discover and monitor them.

Some asteroids may have even been comets once. Comets lose most of the volatiles during passages
through the inner inner Solar System, leaving behind a dark, low-albedo nuclei. For example, asteroids 
944~Hidalgo and 2201~Oljato are believed to move in an elliptical comet-like orbit 
(e.g., Degewij \& Tedesco 1982). Asteroid 2201~Oljato also has an unusual UV reflectance, 
which has been interpreted as Rayleigh scattering from a coma around it (McFadden et al. 1984; 
Rickman 1985). The nature of their orbits and their meteor-shower associations are also in favour of
this hypothesis (see Weissman et al. 2002 for extensive discussion).  
 
\subsubsection{Earth-Moon Trojans}

As one of the new and a more speculative proposals, we will also try to look for the Earth trojans and
Moon trojans. Trojans are asteroids that co-rotate with a planet librating around its $L_4$ or
$L_5$ Lagrangian points of stability (trojan points). For example, Jupiter trojans are about as 
numerous as the asteroids of the asteroid belt (Yoshida \& Nakamura 2005). Other planets' trojans: 
Mars, Saturn, Uranus and Neptune were also discovered. Earth trojans are those that co-orbit 
the Sun with the Earth, librating around Earth--Sun Lagrangian points $L_4$ (leading $60^{\circ}$) 
and $L_5$ (trailing $60^{\circ}$). In 2011, asteroid 2010 TK7 was confirmed 
to be the first discovered Earth trojan. It is located in the $L_4$ (Connors et al. 2011). Earth trojans are 
difficult to detect from the ground as they are generally low in the sky, either just after the sunset or 
just before the sunrise, as seen from Earth. In addition to planetary trojans, there can be moon trojans, 
which are small bodies residing at the trojan point of a planetary moon. For example, Saturnian moons 
Dione and Tethys have trojan asteroids, two each. Though Saturnian moons are much smaller than the 
Earth's Moon (their masses are $1/10^{4}$ of the Earth's), no Moon's trojans were discovered. 
Looking at the trojan points from the Moon is an interesting opportunity to detect these elusive asteroids.

\subsection{UV transient sky}

Astrophysical transient events include bursts, flashes, flares, SNe, 
gamma-ray bursts (GRB), etc. The rate of discoveries of variable objects is constantly
increasing with larger telescopes and newer techniques coming on line, for example,
the Palomar Transient Factory (PTF) alone discovers up to 12 SNe per night. The 
last decade has been the witness to the importance 
of the UV space missions in detecting transients in UV with GALEX discovery of UV 
flares from flare stars and from disruption of a star by a supermassive black hole (SMBH). 
Both {\it SWIFT} and GALEX have been yielding impressive results (e.g. Botticella et al. 2010, 
Rabinak \& Waxman 2011). In fact, the discovery rate of UV transients could only 
increase by several orders of magnitude if a wide-field UV instrument is launched 
(Gezari et al. 2013).

\subsubsection{SNe and GRBs}

Supernovae, the result of a death of a massive star, are extremely luminous transient events
that can outshine an entire galaxy before fading from view several weeks or even months
later. SN shock breakout is most luminous phase of a core-collapse
explosion, producing UV/X-ray burst of radiation when shock emerges at the surface of the star.
SNe discovered in optical surveys are caught too late, when the UV emission is already fading rapidly.
If we can catch SNe early, 
when the hot, thermal emission from the ejecta is bright in the UV. Schawinski et al. (2008) and 
Gezari et al. (2008) have found, using archival UV data from GALEX, that one supernova (SNLS-04D2dc) 
flared in the UV up to seven hours before the actual explosion suggesting that
the first sign of a supernova is a UV flare. They speculated that the expanding internal shock wave 
rapidly heated the star, shifting its emission into the UV. Given the importance of observations within the 
first microseconds of an explosion, this would be a valuable means of pinpointing stars about to explode.

Gamma-ray bursts (GRBs) are flashes of $\g$-rays associated with extremely energetic explosions in 
distant galaxies. They are the most luminous electromagnetic events known to occur in the Universe. 
Bursts can last from milliseconds to several minutes, although a typical burst lasts a few seconds. 
The initial burst is usually followed by a longer-lived "afterglow" emitted at longer wavelengths 
(X-ray, UV, optical, IR and radio). Most observed GRBs are believed to be produced by a narrow beam of intense 
radiation released during a supernova event, as a rapidly rotating, high-mass star collapses to form 
a black hole and called long-GRB. A subclass of GRBs, the short-GRBs appear to originate from a 
different process, possibly the merger of binary neutron stars. The progenitors of long-duration 
GRBs are hypothesized to be the Wolf-Rayet (WR) stars.

\subsubsection{Flaring stars}

A vast majority of UV short-period transients is associated with a single physical type of
an astrophysical source, that is, flare eruptions of $K$ and $M$ stars. M dwarfs account 
for more than $75\%$ of stellar population within 1 kpc of the Sun. The subset
of these stars known as flare stars possess extremely active chromospheres and coronae.
The first evidence of flare activity is best observed in the UV continuum. The rise time
is sharper in the late $M$ dwarfs, a few seconds, reaching few minutes in the $K$ or
early $M$ dwarfs. These flares exceed the quiescent state luminosity by 2 or 3 orders of magnitude.
The peak temperature of the flare reaches $10^3$--$3\times 10^4$ K, implying a UV peak with
luminosities of $10^{25}$--$10^{26}$ W. For such a flare on, say Proxima Cen, this would
correspond to a flux at Earth of $10^5-10^6$ UV photons per cm$^2$ for periods of
several hours. The $M$ dwarf flare 100 pc away will produce UV flux at Earth of $10$--$10^2$ 
phots/cm$^2$.   

Gezari et al. (2013) estimate the surface density of $M$ dwarf transients as 15/deg$^2$/yr, which for a 
$\sim 7$ deg$^2$ telescope would result in annual rate of $10^5$/yr, or 4 per lunar day. Kulkarni \& Rau 
(2006) concluded that the fast transient (which change on timescales of an hour or less) annual rate
in B band of $10^8$/yr, or three events per second, detected in the Deep Lens Survey, is entirely due to the 
flares from Galactic M dwarfs. Using this annual rate $R_f^{\rm dMe} \approx 10^8$/yr, we obtain for LUCI FOV 
of $\D\Omega=2.15\times 10^{-3}$ sr and for $\tau \sim 5\,\text{hrs}\approx 20,000$ sec exposure, the number
of detected flares 
\be
N_f=\fr{R_f^{\rm dMe} \tau \D\Omega}{4\pi} \approx 11\,,
\ee
Since flaring time scales range from minutes to hours, individual flares will be 
observed only once during their time evolution. Therefore, each epoch of each
star must be analyzed for a flare signature. However, if LUCI detects a flare in real time and there is
immediate transmission, the follow-up observations from the ground can be performed at once.

There are many variable stars that are bright in the UV. For example, eclipsing binaries (EBs) 
with $O$- to late $B$-type stars have significant UV fluxes. Extending the wavelength baseline 
to the FUV vastly improves the determination of $T_{\rm eff}$ and ISM absorption. 
Another type, the WR stars, are particular interesting as they are hypothesized to be  
the progenitors of long-duration GRBs. WR stars are evolved, massive, extremely hot and very 
luminous stars, which are losing mass rapidly through a very strong stellar wind. They are 
extremely rare because they are massive and thus short-lived.  

\subsubsection{Tidal disruption events}

There is an increasing observational evidence
for a series of UV/X-ray flares from stars getting disrupted by massive black holes,
initially from a star being blown apart by an internal nuclear explosion, subsequently
from the debris falling into the black hole and finally from the debris stream collision
with itself \citep{Gezari2012}. An intermediate-mass black hole (IMBH) is
capable of completely destroying a white dwarf (WD) that ventures too close to it, 
releasing up to $10^{48}$ ergs peaking in EUV. In fact, observations of daily 
high-energy flares at the Galactic centre led to a conclusion that the central SMBH
disrupts asteroids larger than 10 km in size on a daily basis, producing short-duration
flares with the maximum possible luminosity of the order $10^{39}$ erg/sec (Zubovas et al. 2012), 
and occasionally a planet is disrupted resulting in
longer and brighter flare. This was seen in the galaxy NGC 4845, where a 14--30 Jupiter-size 
object was disrupted by a central black hole (Niko{\l}ajuk \& Walter 2013).

\subsubsection{UV transients from celestical collisions}

Collisions are common on all astronomical scales, from asteroids bombarding planets to 
massive galaxy clusters passing through each other (for ex., Jee et al. 2007). Collisions on all 
scales can be very energetic. Here we consider the collisions that occur in planetary systems.

\paragraph{Asteroidal collisions}

\begin{figure}[h!]
\includegraphics[scale=0.37]{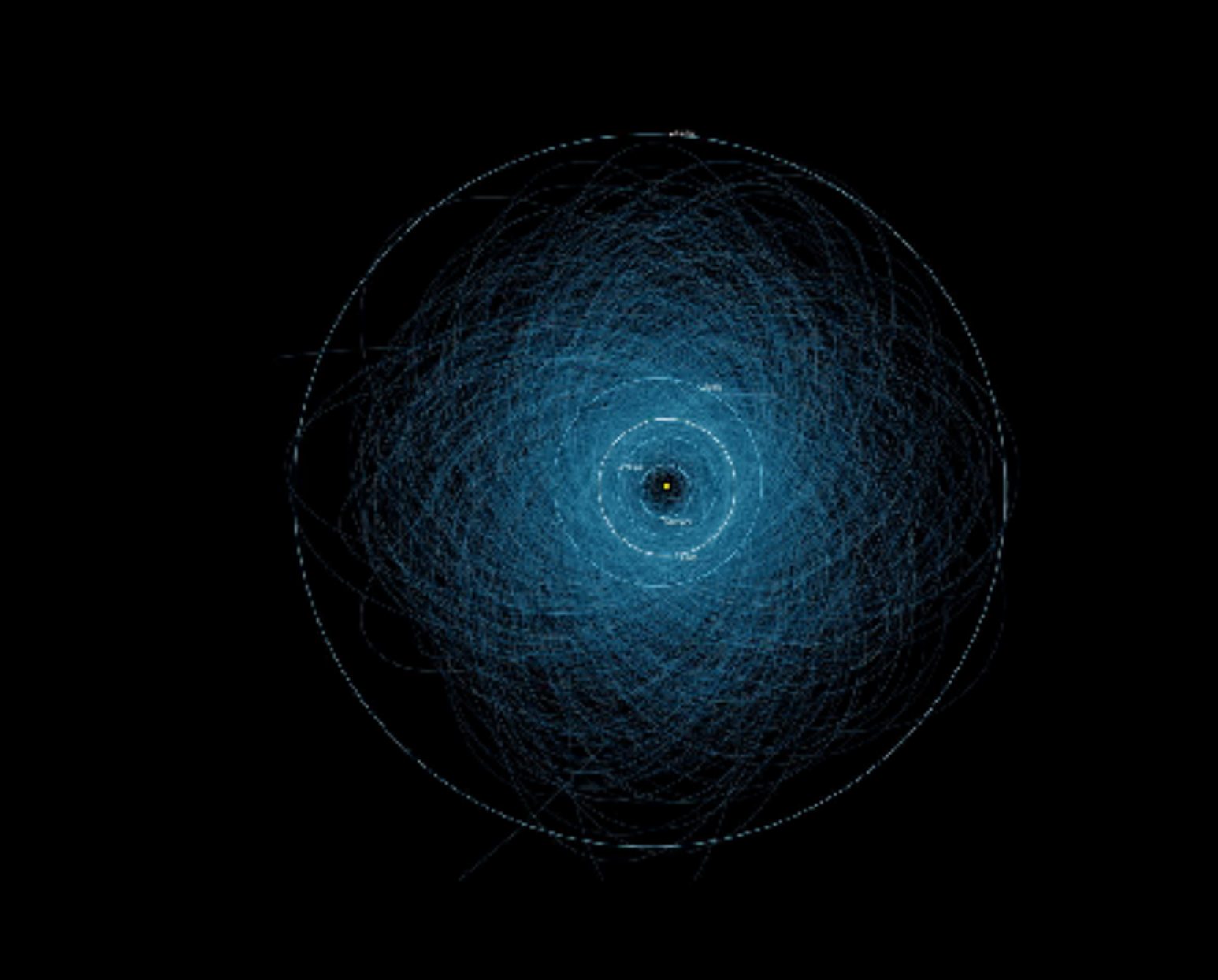}
\includegraphics[scale=0.38]{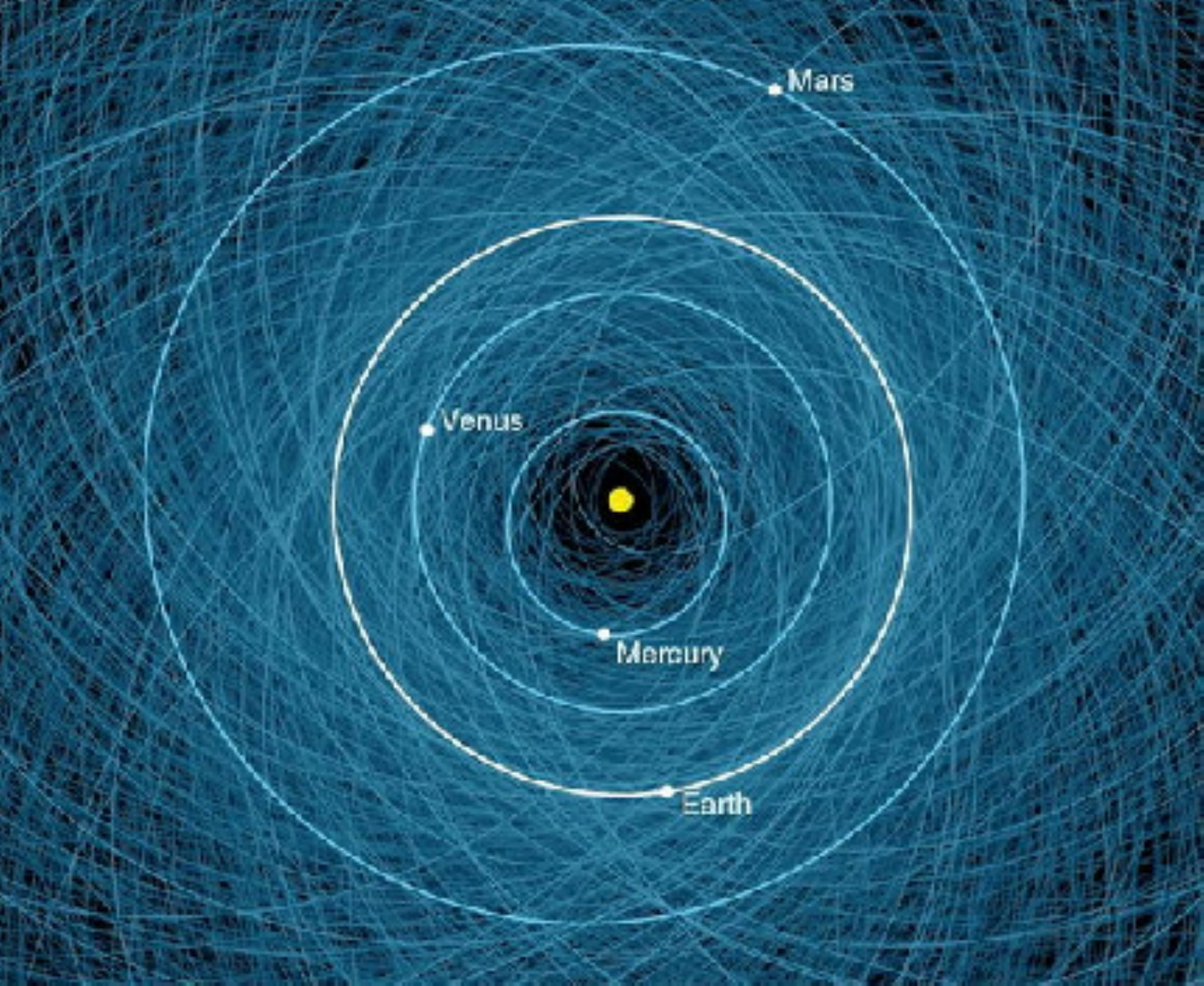}
\caption{These NASA maps show the orbits of over 1,400 known PHA, as of early 
2013. Source: NASA Planetary Photojournal.}
\label{fig:PHA}
\end{figure}

Though many PHA orbits are mapped (Fig.~\ref{fig:PHA}) and none are known 
to be on a collision course with the Earth in the next century, asteroidal 
collisions may change their orbits and take them off the plotted/predicted course. 
In addition to the deviating influence of larger planets like Jupiter, astronomers 
estimate that modest-sized asteroids in the main belt collide with each other 
about once a year, producing fragments that scatter and may unexpectedly 
crash on Earth.

Detecting asteroidal collision in progress is tricky because large impacts are rare, while
 small ones are faint and largely go unnoticed, therefore we detect only the aftermaths. 
Recently two unusual objects were discovered which initially were designated as comets due to
the dust trails,  P/2010 A2 (LINEAR) and P/2013 P5. The comet-like tail of P/2010 A2
was noticed by the Lincoln Near-Earth Research (LINEAR) Program Sky Survey in January 2010. 
Only HST's deep imaging of the nucleus revealed a point-like object with a criss-crossing 
dust feature, which led to a conclusion that this comet-like appearence is actually a result 
of an impact that happened just few weeks previously (Jewitt et al. 2010). The smaller 
impactor was most probably vaporized and material was stripped 
from the larger one. The X-like dust pattern at the head of the tail was most probably the result of an 
asymmetric collision and ejection of material. Both colliding asteroids were too
small and would have gone undetected but for the impact, which was as powerful as a small atomic
bomb (Harrington \& Villard 2010). P/2010 A2 itself is consistent with membership in the Flora 
asteroid family, produced by collisional shattering of a parent body
millions of years ago. One fragment of this family could even have been responsible for the K-T event. 
Another member of this family, asteroid P/2013 P5, also displays unusual 6-tail comet-like
behaviour and could be a fragment from an asteroid collision that occurred roughly 200 million years ago
(Jewitt et al. 2013). 

Subsequently, SWIFT observed the asteroid 596 Scheila, which brightened unexpectedly in the end 
of 2010 after probably colliding with a much smaller asteroid (Moreno et al. 2011). 
596 Scheila is approximately 113 km 
across and orbits the Sun in the Main Belt with a period of $\sim$five years. It is displaying nearly 
the same pattern as P/2010 A2; it was noticed just weeks afer the collision, best explained by a 
4.9 km/sec impact by a smaller 30-m asteroid. Such asteroids are colliding with a frequency of 
about one in five years in the Main Belt (Bodewits et al. 2011).

\paragraph{Collisions in exo-planetary systems}

In a standard model of planetary system formation, the final stage is now strongly 
associated with a significant number of giant impacts on each of the young forming 
planets (Stern 1994). The Mars-sized terrestrial impactor, believed to be responsible 
for the formation of the Moon, was such an impact (Stevenson 1987; Canup \& Asphaug 2001), 
and Mars’ primitive atmosphere is believed to have been lost through a similar giant impact 
catastrophic event 4 Ga ago (first proposed by Hancock, Bauval and Grigsby (1998), 
Spexarth (2004), and suggested by data returned by the Curiosity rover, see e.g., 
Webster et al. 2013). Even a mature Solar Systems, like ours, experience large-scale collisions, 
where even planets can collide and be destroyed in the process. The latest example is of the few Gyr old
star BD+20307 \citep{{Song2005},{Weinberger2008}}, where excess of warm
dust indicates that two terrestrial-size planets have collided and destroyed each other. 

Most known exoplanets orbit solar-type stars, i.e. MS stars of $F$, $G$, $K$ and $M$
categories, mostly within 600 pc from our Sun (e.g. Shchekinov et al. 2013). Collision
of a Jupiter planet with a terrestrial size planet results in a UV flare in 100--300 nm range
that outshines at flare peak time any MS host star later than $B$5 by orders of magnitude.
For example, $\tau$ Ceti has 10 times more dust than our Sun. Any planet in its system 
would suffer from large impact events roughly ten times more frequently than the Earth
(Greaves et al. 2004). At 150 nm, the UV flux from a star like $\tau$ Ceti ($G$8V type) 
at 10 pc is $\sim 1\times 10^{-14}$ erg/cm$^2$/sec, while the UV flare may produce 
$\sim 4\times 10^{-10}$ erg/cm$^2$/sec. The flux contrast is dramatic. If a planet 
(a hot Jupiter, for example) impacts a host star, the flare with the peak temperature of 
$\sim 10^5$ K would generate a soft X-ray flash lasting 
for few hours, followed by the FUV radiation for few days, with 
flux at Earth of $0.3$ erg/cm$^2$/sec at a 10 pc distance (equivalent to $\sim 2.3\times 
10^{10}$ phots/cm$^2$/sec at 150 nm). Collisions of two hot Jupiters would produce a flare lasting 
for hours with peak $T\sim 2\times 10^5$. For an event duration of 10 hours, the FUV fluence
at Earth from collision at 10 pc would be $\sim 10^5$ phots/cm$^2$, while the host star
itself would most probably constitute only $\sim 1$ phots/cm$^2$ (e.g. Safonova et al. 2007). 

To estimate the annual rate of the events, we consider a young planetary system, where the
central star is surrounded by planetesimals of different sizes. Within a volume of (1 A.U.)$^3$ there
can be $10^8-10^9$ planetisimals, where few can be as massive as $10-100\, M_{\oplus}$ 
with sizes 100 km dia and larger. For example, it takes $10^6$ embryos of 100 km diameter to make 
up one planet of 10,000 km. At 
typical velocities of 100 km/sec, a typical system may have 10 collisions per year. If there are
billions of planetary systems in our Galaxy (for ex. Cassan et al. 2013) and more 
than a third of them are young, $< 3$ Gyr old (Safonova et al. 2014), there can be 
$10^8$ such systems in a Galaxy and we can expect $10^9$ such events per year. This 
number may actually even represent only a lower limit. In LUCI FOV of $\D\Omega = 2.15\times 
10^{-3}$, the number of events per exposure, that can last up to $\tau=20,000$ sec, 
can be as high as 
\be
N_f=\fr{R_f^{\rm pl} \tau \D\Omega}{4\pi} \approx 114\,,
\ee
if we look at the Galactic plane. 

Due to the temporal modulation (Safonova et al. 2007), these events may be distinguished 
from the dMe flares. Occurence of repeated flashes of diminishing intensity due to the infall
and subsequent rebound of debris falling back on the impactor is an additional clue. Every 
such collision will be followed by a long-duration IR afterglow 
that can be detected subsequently on the ground as a follow-up.

\section{Summary}
 
Despite being a small and experimental payload, LUCI will have an advantage 
of a wide FOV ($\sim 7$ deg$^2$) over the few instruments that previously operated in UV.  In 
Table~\ref{table:parameters} we compare the general mission parameters for
GALEX, UVIT and LUCI. We have calculated the point source sensitivity (PSS),
the reciprocal of the integration time needed to reach a given S/N, for a perfect 
telescope (Klimas et al. 2010),
\be
S_0=\fr{\left(3\pi^2 -16\right)d^4}{48\lambda^2}\,,
\label{eq:PSS}
\ee
where $d$ is the aperture diameter and $\lambda$ is the effective (operating) wavelength. 
In addition, we have calculated the survey parameter for comparing the ability of UV 
missions to estimate the UV background (Henry 1982; Brosch 1999),
\be
S=\fr{100}{A\D\O \D\lambda}\,,
\label{eq:surveypar}
\ee
where $A$ is the collecting aperture in cm$^2$, $\D\O$ is the FOV in sr, and $\D\lambda$ 
is the bandpass. 

\begin{table}[ht!]
\caption{Comparison of mission parameters in FUV channels for equivalent ideal telescopes based 
on Eqs.~\ref{eq:PSS} and \ref{eq:surveypar}.}
\vskip 0.1in
\begin{center}
\label{table:parameters}
\begin{tabular}{lccc}
\hline
 Mission parameter        &GALEX     &   UVIT               &LUCI                       \\\hline
A   (cm$^2$)              & 1950       &      880                  &706                           \\ 
No. of pixels               &4K$\times 4$K & $512\times 512$  &2K$\times 2$K    \\  
Pixel scale (arcsec/pix)  &1.5        &  3                     &$\sim 7$   \\ 
FOV (deg$^2$)           & 1.1        &  0.7                  & $\sim 7$                       \\ 
FOV (sr)                    &$3.35\times 10^{-4}$&$2.13\times 10^{-4}$& $2.15\times 10^{-3}$ \\ 
FUV bandpass (nm)      &  40         &      50               &   50                               \\
$\lambda^{\dag}$ (nm)& 151.6    & $151. 4$             & assume 150.0            \\
Relative PSS                &   1         &   $\sim 0.3$      &        0.13                   \\ 
S (1/cm$^3$)             &$3.06\times 10^7$&$1.34\times 10^8$ &$1.32\times 10^7$   \\
\hline
\end{tabular}
\end{center}
$^{\dag}$\footnotesize{GALEX effective wavelength for FUV channel is from Hunter et al. (2010).
The effective wavelengths for UVIT FUV filters were estimated in Ravichandran et al. (2013)}.
\end{table}

In this paper we have shown that in spite of being a small and experimental instrument, LUCI is a viable 
preliminary design for a lunar UV telescope that would be able to provide some unique scientific results. 
It, and similar small payloads, may pave the way to UV science in a period of dwindling space science budgets.

\begin{acknowledgements}

MS is thankful to Bernard Foing for his valuable comments and suggestions on the possible scientific areas 
we can address with LUCI. We are also thankful to the anonymous referee for very helpful comments and
suggestions.

\end{acknowledgements}

\end{document}